\documentclass[letter,longauth]{aa} 

\usepackage{graphicx}
\usepackage{lscape}
\usepackage{txfonts}
\usepackage{caption}
\usepackage{color}
\usepackage{grffile}
\usepackage{afterpage}
\usepackage{amsmath}
\usepackage{natbib}
\newcommand{\uchii}{UC-H\,{\scriptsize II}}

\newcommand{\kms}{km~s$^{-1}$}

\newcommand{\msol}{M$_{\odot}$}
\newcommand{\lsol}{L$_{\odot}$}

\begin{document}

   \title{ALMA survey of massive cluster progenitors  from ATLASGAL}
   \subtitle{Limited fragmentation at the early evolutionary stage of massive clumps}

   \author{T. Csengeri
          \inst{1}
         \and
           S. Bontemps
          \inst{2}
         \and
           F. Wyrowski
          \inst{1}
          \and
          F. Motte
          \inst{3,4}
        \and
         K. M. Menten
      \inst{1}
         \and
         H. Beuther
         \inst{5}
         \and 
         L. Bronfman
          \inst{6}
        \and {B. Commer\c con}          \inst{7}
        \and {E. Chapillon}          \inst{2,8}
        \and{A. Duarte-Cabral}            \inst{9,13}
        \and {G. A. Fuller}          \inst{10}
        \and{Th. Henning}          \inst{5}
        \and{S. Leurini}           \inst{1}
       \and{S. Longmore}   \inst{11}
       \and {A. Palau}    \inst{12}
      \and {N. Peretto}    \inst{13}
       \and{F. Schuller}  \inst{1}
       \and{J.\,C. Tan}  \inst{14}
       \and{L. Testi}      \inst{15,16}
      \and{A. Traficante} \inst{17}
       \and{J. S. Urquhart} \inst{18}
         }

   \institute{Max Planck Institute for Radioastronomy,
              Auf dem H\"ugel 69, 53121 Bonn, Germany
              \email{csengeri@mpifr-bonn.mpg.de}
        \and
          OASU/LAB-UMR5804, CNRS, Universit\'e Bordeaux, all\'ee Geoffroy Saint-Hilaire, 33615 Pessac, France 
        \and
         Institut de Plan\' etologie et d'Astrophysique de Grenoble, Univ. Grenoble Alpes -- CNRS-INSU, BP 53, 38041 Grenoble Cedex 9, France
        \and
       Laboratoire AIM Paris Saclay, CEA-INSU/CNRS-Universit\'e Paris Diderot, IRFU/SAp CEA-Saclay, 91191 Gif-sur-Yvette, France
           \and
          Max Planck Institute for Astronomy, K\"onigstuhl 17, 69117 Heidelberg, Germany
          \and
            Departamento de Astronom\'{i}a, Universidad de Chile, Casilla 36-D, Santiago, Chile
          \and
          Univ. Lyon, ENS de Lyon, Univ Lyon1, CNRS, Centre de Recherche Astrophysique de Lyon UMR5574, F-69007, Lyon, France
          \and{IRAM, 300 rue de la piscine, 38406, Saint-Martin-d'H\'eres, France}
           \and{School of Physics and Astronomy, University of Exeter}
          \and
           {Jodrell Bank Centre for Astrophysics,
School of Physics and Astronomy, The University of Manchester,
Manchester, M13 9PL, UK}
           \and
          Astrophysics Research Institute, Liverpool John Moores      
               \and
          {Instituto de Radioastronom\'ia y Astrof\'isica, Universidad Nacional Aut\'onoma de M\'exico}
           \and 
           {School of Physics \& Astronomy, Cardiff University}
           \and
          {Departments of Astronomy and Physics, University of Florida }
             \and
         European Southern Observatory, Karl-Schwarzschild-Strasse 2, D-85748 Garching, Germany
          \and
          INAF-Osservatorio Astrofisico di Arcetri, Largo E. Fermi 5, I-50125 Firenze, Italy
University 
       \and {IAPS-INAF, Via Fosso del Cavaliere, 100, 00133, Rome, Italy}
\and
School of Physical Sciences, University of Kent, 
              Ingram Building, Canterbury, Kent CT2 7NH, UK
           }

   \date{Received , 2016; accepted , 2016}

 
  \abstract
   {The early evolution of massive cluster progenitors is poorly understood.
   We investigate the fragmentation properties from 0.3\,pc to 0.06\,pc scales of a homogenous sample of infrared-quiet massive clumps within 4.5\,kpc selected from the ATLASGAL survey. Using the ALMA 7m array we detect compact dust continuum emission towards all targets, and 
 find that fragmentation, at these scales, is limited. 
 The mass distribution of the fragments uncovers a large fraction of cores above 40\,\msol, corresponding to massive dense cores (MDCs) with masses up to $\sim$400\,\msol. 
 77\% of the clumps contain at most 3 MDCs per clump, and we also reveal single clumps/MDCs. The most massive cores are formed within the more massive clumps, and a high concentration of mass on small scales reveals a high core formation efficiency. The mass of MDCs highly exceeds the local thermal Jeans-mass, and observational evidence is lacking for a sufficiently high level of turbulence  or strong enough magnetic fields to keep the most massive MDCs in equilibrium. If already collapsing, the observed fragmentation properties with a high core formation efficiency are consistent with the collapse setting in at parsec scales.  }


   \maketitle
%

\section{Introduction}

The properties and the evolution of massive clumps hosting the precursors of the highest mass stars currently forming in our Galaxy are poorly known. 
Massive clumps at an early evolutionary phase,
{  thus, prior to the emergence of luminous massive young stellar objects and {\uchii} regions},
are excellent candidates to host high-mass protostars in their earliest stages (e.g.\, \citealp{Zhang2009,Bontemps2010,Csengeri2011a,Csengeri2011b,Palau2013,SM2013}). 
Large samples have only recently been identified based on large area surveys (e.g.\,\citealp{BT2012,Tackenberg2012,Traficante2015,Svoboda2016,Csengeri2017}), which 
show that the early evolutionary stages are short lived (e.g.\,\citealp{M07,Csengeri2014}), as star formation proceeds rapidly. 
Using the Atacama Large Millimeter/submillimeter Array (ALMA), 
here we present the first results of a statistical study of early stage fragmentation 
to shed light on the physical processes 
at the origin of high-mass collapsing entities, 
and to search for the youngest
 precursors of O-type stars.

\section{The sample of infrared quiet massive clumps}

Based on a flux limited sample  
of the 870 $\mu$m APEX Telescope LArge Survey of the GAlaxy (ATLASGAL, \citealp{Schuller2009, Csengeri2014}), \citet{Csengeri2017}
 identified the complete sample of massive infrared quiet clumps 
with { the highest peak surface density ($\Sigma_{\rm cl}\geq0.5$\,g\,cm$^{-2}$)\footnote{In the ATLASGAL beam of {19{\rlap{\arcsec}{.}2}.}}
and low bolometric luminosity, $L_{\rm bol}$$<$$10^4$\,\lsol,
corresponding to the ZAMS luminosity of a late O type star.
Their large mass reservoir and low luminosity suggest
that infrared quiet massive clumps correspond to the early evolutionary phase,  
some already exhibiting signs of ongoing (high-mass) 
star formation such as EGOs and Class~II methanol masers. 
Here we present the sample of  
35 infrared quiet massive clumps 
located within $d\leq4.5$\,kpc, which could be conveniently grouped on the sky as targets for ALMA. 
They cover 70\% of all the most massive and nearby infrared quiet clumps from 
\citet{Csengeri2017}, and 
are thus a representative selection of a 
 homogenous sample of
 early phase massive clumps in the inner Galaxy.
\section{Observations and data reduction}

We present observations carried out in Cycle~2 with the ALMA 7m array using 
$9$ to $11$ of the 7m antennas with baselines ranging between 8.2\,m (9.5k$\lambda$) 
to 48.9\,m (53.4k$\lambda$). 
We used a low-resolution wide-band setup
in Band 7, yielding $4\times1.75$\,GHz
effective bandwidth with a spectral resolution of 976.562 kHz.
The four basebands were centred on
347.331, 345.796, 337.061, 335.900\,GHz, respectively.  
The primary beam at this frequency is 28.9\arcsec. 
Each source was observed for $\sim$5.4~min in total.
The system temperature, $T_{\rm sys}$ varies between $100-150$~K.
{  The targets have been
split according to Galactic longitude in five observing groups (Table\,\ref{tab:table-obs}).}

{ 
The data was calibrated using standard procedures in CASA~4.2.1.} 
To obtain line-free continuum images, we first identified the channels with
spectral lines towards each source, and excluding these averaged the remaining channels. 
We used a robust weight of 0.5 for imaging, and the CLEAN algorithm
for the deconvolution, and corrected for the primary beam attenuation. 
The synthesized beam varies between $3.5$\arcsec\ to $4.6$\arcsec\
taking the geometric mean of the major and minor axes. 
{  The noise has been measured in an emission free area close to the center of the maps { including the side-lobes}.} 
The achieved median $rms$ noise level is $54$\,mJy/beam and
varies among the targets due to a combination of restricted bandwidth available for continuum, dynamic range or mediocre observing conditions. In particular for  
groups 4 and 5, the observations have been carried out at low elevation 
resulting in an elongated beam and poor $uv$-sampling.
{  The observing parameters per group are summarized in Table\,\ref{tab:table-obs},
and for each source in Table\,\ref{table:long}.}

\begin{table*}[!ht]
\centering
\caption{Summary of observations.}\label{tab:table-obs}
\begin{tabular}{cccclrrrrrrrrrrr}
\hline\hline
 \multicolumn{2}{c}{Observing group} & Date & Bandpass   & Phase  & Flux  & \multicolumn{3}{c}{Synthesized beam\tablefootmark{a}} & $\sigma_{\rm rms}$ \tablefootmark{b}  \\
& & & calibrator & calibrator & calibrator &  [\arcsec$\times$\arcsec]  &  [$^{\circ}$]& [\arcsec] & [mJy] \\
\hline 
1 & $320<\ell<330^{\circ}$& 8, 16 July 2014& J1427-4206 & J16170-5848 & Titan, Ceres & $5.0\times2.9$& -78.6 & $3.8$ & $19.3-83.5$\\
2 & $330<\ell<340^{\circ}$ & 18, 21 July 2014 &J1427-4206 & J1617-5848 & Titan, Ceres &
 $4.6\times 2.8$&     14.9&      3.6& $20.7-119.2$\\
3 & $340<\ell<350^{\circ}$ & 19, 21 July 2014 & J1517-2422 &  J1636-4102 & Titan, Ceres & 
 $4.7\times 2.6$ &   -83.4 &    3.5 
& $22.9-105.3$ \\
4 & $350<\ell<360^{\circ}$& 14, 15 June 2014 & J1733-1304  & J1717-3342 & Neptune &
$9.2\times 2.4$ &   -76.2 & 4.6 & $28.7-175.8$ \\
5 & $30<\ell<40^{\circ}$& 8 June 2014 & J1751+0939 &J1851+0035 & Neptune &  
 $5.8\times2.4$&   -68.2 & 3.7 & $16.4-45.8$ \\
\hline
\end{tabular}
 \tablefoot{ 
\tablefoottext{a}{Averaged properties.}
\tablefoottext{b}{The minimum and maximum $\sigma_{\rm rms}$ noise is averaged over the line-free channels in the total 7.5\,GHz bandwidth.}
}\end{table*}
   \begin{figure}[]
   \centering
   \includegraphics[width=0.74\linewidth]{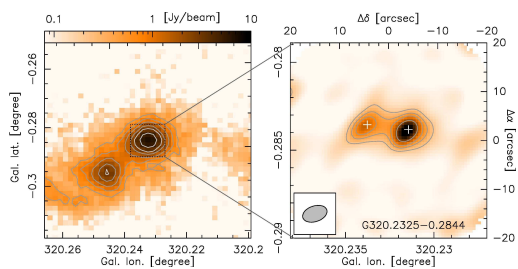}
      \caption{\small
   		 {\it Left:}  Clump-scale view by ATLASGAL of an example source.
		 {\it Right:}  Line-free continuum emission at 345\,GHz by the ALMA 7m array. Contours start at 7$\sigma_{\rm rms}$ noise and increase in a logarithmic scale. White crosses mark the extracted sources (see Table\,\ref{table:long}). The synthesized beam is shown in the lower left corner. 
		}
              \label{fig:overview}%
    \end{figure}

\longtab[1]{
\begin{landscape}
\begin{longtable}{rcccccccccccccccccc}
\caption{Summary of physical properties of the sample. \label{table:long}}\\
\hline \hline
Source & \multicolumn{2}{c}{Position} & $F_{\nu}$ & $S_{\nu}$ & $\Theta_{\rm A}$ & $\Theta_{\rm B}$ & beam & $FWHM$ & $d$ & $M_{\rm core}$ & $\Sigma_{\rm core}$ \\
& [RA J2000] & [DEC 2000] & [Jy/beam] & [Jy] & [\arcsec] & [\arcsec] & [\arcsec] & [\arcsec] & [kpc] & [\msol] & [g\,cm$^{-2}$] &  &  \\ 
\hline
\endfirsthead
\caption{continued.}\\
\hline\hline
Source & \multicolumn{2}{c}{Position} & $F_{\nu}$ & $S_{\nu}$ & $\Theta_{\rm A}$ & $\Theta_{\rm B}$ & beam & $FWHM$ & $d$ & $M_{\rm core}$ & $\Sigma_{\rm core}$  \\
& [RA J2000] & [DEC 2000] & [Jy/beam] & [Jy] & [\arcsec] & [\arcsec] & [\arcsec] & [\arcsec] & [kpc] & [\msol] & [g\,cm$^{-2}$] & &  \\
 \hline
\endhead
\hline
\endfoot
320.2325-0.2844-MM1  &   227.46639  &   -58.42730  &      1.59  &      2.47  &      5.80  &      4.41  &      4.06  &      5.06  &      3.90  &    147.80  &      4.13   \\
               -MM2  &   227.46991  &   -58.42589  &      0.85  &      1.64  &      7.17  &      4.42  &      4.06  &      5.63  &      3.90  &     98.25  &      1.64   \\
321.9348-0.0066-MM1  &   229.93021  &   -57.30158  &      1.24  &      2.65  &      6.47  &      5.18  &      3.97  &      5.79  &      1.80  &     33.68  &      2.27   \\
               -MM2  &   229.92670  &   -57.30223  &      0.75  &      1.48  &      7.17  &      4.35  &      3.97  &      5.58  &      1.80  &     18.84  &      1.46   \\
               -MM3  &   229.92896  &   -57.30329  &      0.32  &      0.61  &      6.98  &      4.33  &      3.97  &      5.50  &      1.80  &      7.74  &      0.64   \\
               -MM4  &   229.92747  &   -57.30112  &      0.28  &      0.37  &      3.79  &      5.60  &      3.97  &      4.61  &      1.80  &      4.76  &      1.04   \\
               -MM5  &   229.92681  &   -57.30411  &      0.21  &      0.31  &      7.17  &      3.18  &      3.97  &      4.77  &      1.80  &      3.93  &      0.67   \\
322.1632+0.6221-MM1  &   229.66623  &   -56.64646  &      1.45  &      2.74  &      6.18  &      4.83  &      3.97  &      5.46  &      3.20  &    110.26  &      2.96   \\
               -MM2  &   229.66615  &   -56.64783  &      0.55  &      1.44  &      9.17  &      4.53  &      3.97  &      6.45  &      3.20  &     58.07  &      0.85   \\
               -MM3  &   229.66471  &   -56.64973  &      0.39  &      0.73  &      6.02  &      4.88  &      3.97  &      5.42  &      3.20  &     29.33  &      0.81   \\
               -MM4  &   229.66313  &   -56.64825  &      0.26  &      0.29  &      4.68  &      3.68  &      3.97  &      4.15  &      3.20  &     11.51  &      2.93   \\
323.7407-0.2635-MM1  &   232.94024  &   -56.51407  &      2.77  &      5.10  &      6.52  &      4.38  &      3.94  &      5.34  &      2.80  &    157.05  &      5.95   \\
               -MM2  &   232.94057  &   -56.51270  &      1.12  &      1.81  &      7.09  &      3.53  &      3.94  &      5.00  &      2.80  &     55.78  &      2.90   \\
               -MM3  &   232.94111  &   -56.51695  &      0.38  &      1.21  &      6.19  &      7.97  &      3.94  &      7.02  &      2.80  &     37.13  &      0.54   \\
326.4745+0.7027-MM1  &   235.81914  &   -54.12054  &      3.41  &      6.10  &      5.77  &      4.76  &      3.92  &      5.24  &      2.50  &    149.75  &      7.66   \\
               -MM2  &   235.81804  &   -54.12175  &      0.73  &      1.10  &      3.92  &      5.91  &      3.92  &      4.81  &      2.50  &     26.92  &      2.13   \\
               -MM3  &   235.82181  &   -54.12051  &      0.52  &      0.81  &      4.15  &      5.72  &      3.92  &      4.87  &      2.50  &     19.77  &      1.46   \\
326.6411+0.6127-MM1  &   236.13754  &   -54.09115  &      1.85  &      4.42  &      5.09  &      7.10  &      3.89  &      6.01  &      2.50  &    108.45  &      3.20   \\
               -MM2  &   236.14062  &   -54.08943  &      0.31  &      0.55  &      5.41  &      4.99  &      3.89  &      5.20  &      2.50  &     13.62  &      0.71   \\
326.6706+0.5539-MM1  &   236.23851  &   -54.12047  &      1.44  &      2.40  &      5.86  &      4.27  &      3.88  &      5.00  &      2.50  &     58.85  &      3.66   \\
               -MM2  &   236.23868  &   -54.11912  &      0.53  &      1.68  &     10.16  &      4.70  &      3.88  &      6.91  &      2.50  &     41.33  &      0.78   \\
               -MM3  &   236.23985  &   -54.11788  &      0.28  &      0.56  &      3.91  &      7.72  &      3.88  &      5.49  &      2.50  &     13.72  &      0.56   \\
328.2353-0.5481-MM1  &   239.49290  &   -53.98994  &      1.99  &      3.83  &      5.43  &      5.18  &      3.82  &      5.30  &      2.50  &     94.10  &      4.31   \\
               -MM2  &   239.49205  &   -53.99156  &      0.63  &      1.36  &      5.28  &      5.98  &      3.82  &      5.62  &      2.50  &     33.30  &      1.22   \\
               -MM3  &   239.49412  &   -53.98866  &      0.38  &      0.60  &      3.52  &      6.49  &      3.82  &      4.78  &      2.50  &     14.69  &      1.10   \\
328.2551-0.5321-MM1  &   239.49919  &   -53.96676  &      2.40  &      3.82  &      6.09  &      3.85  &      3.84  &      4.84  &      2.50  &     93.70  &      6.67   \\
               -MM2  &   239.50159  &   -53.96471  &      0.41  &      1.38  &      7.90  &      6.25  &      3.84  &      7.03  &      2.50  &     33.87  &      0.61   \\
               -MM3  &   239.49997  &   -53.96570  &      0.28  &      0.29  &      2.99  &      5.18  &      3.84  &      3.94  &      2.50  &      7.20  &      5.91   \\
329.0303-0.2022-MM1  &   240.12633  &   -53.20764  &      1.80  &      2.70  &      5.90  &      3.68  &      3.80  &      4.66  &      2.50  &     66.38  &      5.65   \\
               -MM2  &   240.13261  &   -53.21375  &      1.76  &      2.89  &      5.70  &      4.15  &      3.80  &      4.86  &      2.50  &     70.86  &      4.76   \\
               -MM3  &   240.13267  &   -53.21539  &      0.63  &      1.28  &      6.18  &      4.75  &      3.80  &      5.42  &      2.50  &     31.45  &      1.31   \\
               -MM4  &   240.12675  &   -53.21026  &      0.39  &      0.55  &      3.90  &      5.20  &      3.80  &      4.50  &      2.50  &     13.49  &      1.43   \\
               -MM5  &   240.13436  &   -53.21298  &      0.38  &      0.78  &      3.83  &      7.76  &      3.80  &      5.45  &      2.50  &     19.14  &      0.78   \\
               -MM6\tablefootmark{$\dagger$}  &   240.12874  &   -53.20814  &      0.24  &      0.22  &      3.50  &      3.83  &      3.80  &      3.80  &      2.50  &      5.44  &      0.23   \\
329.1835-0.3147-MM1  &   240.44578  &   -53.19542  &      2.19  &      3.63  &      6.06  &      4.00  &      3.83  &      4.92  &      4.20  &    251.79  &      5.76   \\
332.9630-0.6781-MM1  &   245.34527  &   -50.88301  &      1.37  &      2.50  &      5.75  &      4.30  &      3.68  &      4.97  &      4.20  &    172.85  &      3.41   \\
               -MM2  &   245.34521  &   -50.88081  &      0.29  &      0.79  &      5.46  &      6.69  &      3.68  &      6.04  &      4.20  &     54.60  &      0.52   \\
333.1298-0.5602-MM1  &   245.40112  &   -50.67972  &      1.26  &      2.92  &      7.08  &      4.43  &      3.68  &      5.60  &      4.20  &    201.96  &      2.49   \\
               -MM2  &   245.39715  &   -50.68238  &      0.73  &      1.19  &      5.30  &      4.14  &      3.68  &      4.68  &      4.20  &     82.42  &      2.15   \\
               -MM3  &   245.39899  &   -50.68089  &      0.57  &      0.86  &      6.02  &      3.42  &      3.68  &      4.54  &      4.20  &     59.80  &      1.86   \\
               -MM4  &   245.39882  &   -50.67936  &      0.21  &      0.26  &      4.97  &      3.34  &      3.68  &      4.07  &      4.20  &     17.79  &      1.27   \\
333.4659-0.1641-MM1  &   245.33441  &   -50.16293  &      2.26  &      3.69  &      5.64  &      3.89  &      3.67  &      4.68  &      4.20  &    255.59  &      6.62   \\
               -MM2  &   245.33505  &   -50.16183  &      0.64  &      1.23  &      8.01  &      3.23  &      3.67  &      5.09  &      4.20  &     84.98  &      1.50   \\
               -MM3  &   245.33583  &   -50.16502  &      0.46  &      0.54  &      4.91  &      3.21  &      3.67  &      3.97  &      4.20  &     37.19  &      3.54   \\
               -MM4  &   245.33426  &   -50.16054  &      0.29  &      0.66  &      6.62  &      4.64  &      3.67  &      5.54  &      4.20  &     45.74  &      0.58   \\
335.5857-0.2906-MM1  &   247.74493  &   -48.73158  &      4.53  &      7.11  &      5.64  &      3.91  &      3.75  &      4.70  &      3.80  &    403.36  &     13.52   \\
335.7896+0.1737-MM1  &   247.44705  &   -48.26453  &      2.05  &      3.68  &      6.42  &      3.78  &      3.68  &      4.93  &      3.80  &    208.45  &      5.20   \\
               -MM2  &   247.44198  &   -48.26379  &      0.94  &      1.29  &      5.67  &      3.29  &      3.68  &      4.32  &      3.80  &     73.42  &      3.83   \\
               -MM3  &   247.44639  &   -48.26336  &      0.57  &      1.38  &      7.89  &      4.13  &      3.68  &      5.71  &      3.80  &     78.23  &      1.10   \\
336.0177-0.8283-MM1  &   248.78877  &   -48.78004  &      2.73  &      6.23  &      6.38  &      4.76  &      3.65  &      5.51  &      3.80  &    353.15  &      5.56   \\
               -MM2  &   248.78620  &   -48.77915  &      0.84  &      1.38  &      5.80  &      3.78  &      3.65  &      4.68  &      3.80  &     78.32  &      2.44   \\
336.2884-1.2547-MM1  &   249.53804  &   -48.86564  &      1.23  &      1.67  &      5.40  &      3.29  &      3.62  &      4.21  &      1.80  &     21.28  &      5.47   \\
               -MM2  &   249.54189  &   -48.86684  &      0.32  &      0.52  &      5.24  &      4.05  &      3.62  &      4.61  &      1.80  &      6.60  &      0.97   \\
               -MM3  &   249.54265  &   -48.87053  &      0.39  &      0.53  &      4.78  &      3.71  &      3.62  &      4.21  &      1.80  &      6.74  &      1.74   \\
338.9249+0.5539-MM1  &   250.14232  &   -45.69342  &      2.53  &      4.11  &      6.22  &      3.40  &      3.61  &      4.60  &      3.90  &    245.59  &      7.70   \\
               -MM2  &   250.13965  &   -45.69364  &      1.38  &      3.31  &      5.93  &      5.27  &      3.61  &      5.59  &      3.90  &    197.56  &      2.76   \\
               -MM3  &   250.14142  &   -45.69537  &      0.59  &      0.98  &      6.00  &      3.60  &      3.61  &      4.65  &      3.90  &     58.46  &      1.74   \\
338.9266+0.6329-MM1  &   250.05819  &   -45.64154  &      1.21  &      2.26  &      7.06  &      3.49  &      3.62  &      4.96  &      3.90  &    135.19  &      2.99   \\
               -MM2  &   250.05655  &   -45.64261  &      0.69  &      1.18  &      6.40  &      3.54  &      3.62  &      4.76  &      3.90  &     70.76  &      1.89   \\
339.6802-1.2090-MM1  &   252.77450  &   -46.26829  &      1.05  &      2.04  &      5.80  &      4.12  &      3.51  &      4.89  &      1.80  &     25.97  &      2.68   \\
               -MM2  &   252.77592  &   -46.26746  &      0.69  &      1.05  &      5.27  &      3.54  &      3.51  &      4.32  &      1.80  &     13.39  &      2.53   \\
               -MM3  &   252.77591  &   -46.26609  &      0.55  &      0.92  &      5.02  &      4.12  &      3.51  &      4.55  &      1.80  &     11.77  &      1.68   \\
340.2740-0.2113-MM1  &   252.22186  &   -45.17281  &      1.73  &      2.99  &      6.48  &      3.27  &      3.50  &      4.60  &      3.80  &    169.40  &      5.09   \\
               -MM2  &   252.22369  &   -45.17239  &      0.61  &      1.09  &      3.52  &      6.24  &      3.50  &      4.69  &      3.80  &     61.83  &      1.71   \\
               -MM3  &   252.22300  &   -45.17360  &      0.42  &      0.66  &      6.53  &      2.92  &      3.50  &      4.37  &      3.80  &     37.15  &      1.46   \\
340.9698-1.0212-MM1  &   253.73937  &   -45.15131  &      2.22  &      3.93  &      5.69  &      3.76  &      3.48  &      4.63  &      1.80  &     49.95  &      6.42   \\
               -MM2  &   253.73742  &   -45.15118  &      0.93  &      1.31  &      4.62  &      3.69  &      3.48  &      4.13  &      1.80  &     16.64  &      4.02   \\
               -MM3  &   253.73480  &   -45.15067  &      0.68  &      1.25  &      6.05  &      3.70  &      3.48  &      4.73  &      1.80  &     15.91  &      1.85   \\
               -MM4  &   253.73435  &   -45.14978  &      0.56  &      0.58  &      4.65  &      2.69  &      3.48  &      3.54  &      1.80  &      7.33  &     20.95   \\
               -MM5  &   253.73785  &   -45.15201  &      0.38  &      0.61  &      6.18  &      3.16  &      3.48  &      4.42  &      1.80  &      7.78  &      1.25   \\
343.1271-0.0632-MM1  &   254.57153  &   -42.86876  &      5.03  &      8.41  &      6.01  &      3.43  &      3.51  &      4.54  &      2.00  &    132.15  &     15.45   \\
               -MM2  &   254.56979  &   -42.86809  &      0.58  &      1.28  &      4.76  &      5.71  &      3.51  &      5.21  &      2.00  &     20.07  &      1.31   \\
343.7559-0.1640-MM1  &   255.20783  &   -42.43592  &      5.26  &      7.35  &      5.08  &      3.22  &      3.42  &      4.04  &      1.80  &     93.53  &     24.01   \\
               -MM2  &   255.20742  &   -42.43678  &      0.81  &      1.29  &      6.65  &      2.80  &      3.42  &      4.32  &      1.80  &     16.37  &      2.83   \\
344.2275-0.5688-MM1  &   256.03213  &   -42.31088  &      7.54  &      9.88  &      5.29  &      3.07  &      3.52  &      4.03  &      2.00  &    155.15  &     39.12   \\
348.1825+0.4829-MM1  &   258.03610  &   -38.51286  &      2.84  &      6.26  &      5.65  &      4.75  &      3.49  &      5.18  &      1.30  &     41.57  &      6.50   \\
               -MM2  &   258.03667  &   -38.51502  &      0.88  &      1.47  &      4.94  &      4.10  &      3.49  &      4.50  &      1.30  &      9.78  &      2.77   \\
               -MM3  &   258.03538  &   -38.51391  &      0.77  &      0.92  &      4.69  &      3.10  &      3.49  &      3.81  &      1.30  &      6.12  &      5.88   \\
348.5493-0.9789-MM1 & 259.83506 & -39.06427 & 1.50 & 3.29 & 5.43 & 4.74 & 3.43 & 5.07 & 1.80 & 41.80 & 3.59  \\
351.4441+0.6579-MM1  &   260.23010  &   -35.75109  &      5.33  &     12.06  &     10.65  &      4.83  &      4.77  &      7.17  &      1.70  &    136.87  &      6.40   \\
               -MM2  &   260.22902  &   -35.75214  &      3.68  &      5.21  &      9.94  &      3.24  &      4.77  &      5.67  &      1.70  &     59.15  &      8.38   \\
               -MM3  &   260.22825  &   -35.75284  &      2.03  &      2.79  &      2.62  &     11.89  &      4.77  &      5.58  &      1.70  &     31.63  &      5.04   \\
               -MM4  &   260.22749  &   -35.75493  &      1.37  &      2.08  &      9.15  &      3.77  &      4.77  &      5.87  &      1.70  &     23.60  &      2.69   \\
354.6154+0.4719-MM1  &   262.57147  &   -33.23189  &      3.63  &      6.16  &      8.46  &      3.63  &      4.26  &      5.54  &      1.70  &     69.86  &      7.45   \\
               -MM2  &   262.57124  &   -33.23081  &      0.97  &      2.60  &     11.53  &      4.23  &      4.26  &      6.98  &      1.70  &     29.50  &      1.29   \\
351.1542+0.7073-MM1  &   259.97866  &   -35.96326  &      0.76  &      2.05  &     11.75  &      5.20  &      4.77  &      7.82  &      1.70  &     23.28  &      0.81   \\
               -MM2  &   259.97969  &   -35.96195  &      0.63  &      0.64  &      6.25  &      3.68  &      4.77  &      4.80  &      1.70  &      7.23  &     37.93   \\
034.2570+0.1656-MM1  &   283.31617  &     1.25434  &      1.23  &      2.16  &      6.94  &      3.56  &      3.75  &      4.97  &      1.56  &     20.62  &      3.09   \\
               -MM2  &   283.31788  &     1.25249  &      0.50  &      0.68  &      2.67  &      7.18  &      3.75  &      4.38  &      1.56  &      6.46  &      2.01   \\
034.4112+0.2344-MM1  &   283.32508  &     1.42376  &      4.41  &      5.97  &      6.30  &      2.98  &      3.72  &      4.33  &      1.56  &     57.06  &     18.50   \\
035.1330-0.7450-MM1  &   284.52579  &     1.61874  &      1.18  &      1.93  &      6.81  &      3.22  &      3.67  &      4.68  &      2.19  &     36.29  &      3.47   \\
               -MM2  &   284.52742  &     1.61736  &      0.30  &      0.46  &      2.86  &      7.13  &      3.67  &      4.52  &      2.19  &      8.69  &      1.01   \\
034.4005+0.2262-MM1  &   283.32802  &     1.41122  &      0.80  &      1.71  &      8.25  &      3.53  &      3.69  &      5.40  &      1.56  &     16.36  &      1.68   \\
               -MM2  &   283.32920  &     1.40900  &      0.49  &      0.81  &      6.76  &      3.37  &      3.69  &      4.77  &      1.56  &      7.75  &      1.35   \\
               -MM3  &   283.32712  &     1.41270  &      0.42  &      0.64  &      6.24  &      3.33  &      3.69  &      4.56  &      1.56  &      6.16  &      1.38   \\
               -MM4\tablefootmark{$\dagger$}  &   283.32969  &     1.40778  &      0.29  &      0.29  &      2.31  &      5.87  &      3.69  &      3.69  &      1.56  &      2.74  &      0.32   \\
               -MM5  &   283.33105  &     1.40900  &      0.25  &      0.27  &      2.29  &      6.49  &      3.69  &      3.86  &      1.56  &      2.62  &      3.43   \\
               -MM6  &   283.32569  &     1.41281  &      0.18  &      0.31  &      3.67  &      6.40  &      3.69  &      4.85  &      1.56  &      2.95  &      0.48   \\
               -MM7\tablefootmark{$\dagger$}  &   283.32869  &     1.40993  &      0.13  &      0.08  &      2.28  &      3.77  &      3.69  &      3.69  &      1.56  &      0.79  &      0.09   \\
               -MM8  &   283.32622  &     1.41144  &      0.14  &      0.19  &      6.35  &      3.00  &      3.69  &      4.36  &      1.56  &      1.81  &      0.53   \\
\hline
\end{longtable}
 \tablefoot{
 \tablefoottext{$\dagger$}{Unresolved sources.}\\
 The full table is available in electronic form at the CDS via anonymous ftp to cdsarc.u-strasbg.fr (130.79.125.5) or via http://cdsweb.u-strasbg.fr/cgi-bin/qcat?J/A\&A/. Column 1 gives the source name, column 2 and 3 lists the position in J2000 equatorial coordinates. Column 4 and 5 give the peak and integrated flux densities, columns 6 and 7 give the FWHM major and minor axes. Column 8 gives the beam size as the geometric mean of the beam major and minor axes. Column 9 gives the beam convolved angular source size. Column 10 gives the distance from \citet{Csengeri2017}. Column 11 and 12 give the core mass and surface density as described in the main text. 
 }
\end{landscape}
}


\section{Results and analysis}\label{sec:res}

Compact continuum emission is detected towards all clumps 
(see Fig.\,\ref{fig:overview} for an example, and
Fig.\,A\,\ref{fig:all} for all targets). 
We find sources that stay single ($\sim$14\%)
at our resolution and sensitivity.
Fragmentation is, in fact, limited towards the majority of the sample;
$45$\% of the clumps hosts up to two, while $77$\% host up to three compact sources.
Only a few clumps host more fragments.

We identify and measure the parameters of the compact sources
using the Gaussclumps task in GILDAS\footnote{Continuum and Line Analysis Single-Dish Software http://www.iram.fr/IRAMFR/GILDAS}, which performs a 2D
Gaussian fitting. A total number of 124 fragments down to a $\sim$7\,$\sigma_{\rm rms}$ noise 
level are systematically identified within the primary beam, where 
the noise is measured towards each field.  
This gives on average, $\bar{N}_{\rm fr}$=3 sources per clump
corresponding to a population of cores 
at the typically achieved physical resolution of $\sim$0.06\,pc.

We can directly compare 
the integrated flux in compact sources seen by the ALMA 7m array with 
the ATLASGAL flux densities measured over the primary beam of the array 
as both datasets have similar centre frequencies\footnote{
The centre frequency for the ALMA dataset is at 341.4\,GHz, while for the 
LABOCA filter, it is around 345\,GHz. 
A spectral index of $-3.5$
gives 10\% change in the flux 
up to a difference of 10\,GHz in the centre frequencies. This is below
our absolute flux uncertainty. 
}.
We recover between 16-47\% 
of the flux, the rest of the emission is filtered above the typically 19\arcsec\
largest angular scale sensitivity 
of the ALMA 7m array observations.

To estimate the mass, we assume optically thin dust emission and use 
the same formula as in \citet{Csengeri2017}; $M = S_{\rm 870\mu m} d^2\,\kappa_{\rm 870\mu m}\,^{-1}\,B_{{\rm 870\mu m}}(T_{\rm d})^{-1}$, where $S_{\rm 870\mu m}$ is the integrated flux density, $d$ is the distance, $\kappa_{\rm 870\mu m}=0.0185$\,g\,cm$^{-2}
$ from \citet{OH1994} accounting for a gas-to-dust ratio of 100, and $B_{\nu}(T_{\rm d})$ is the Planck function. While on the $\sim$0.3\,pc scales of clumps \citet{Csengeri2017} adopt $T_{\rm d}$=18\,K, on the smaller scales of cores heating due to the embedded protostar may result in elevated dust temperatures that are poorly constrained. Following the model of \citet{GoldreichKwan1974}, we estimate 
$T_{\rm d}$=15-38\,K
for the luminosity range of $10^2-10^4$\,\lsol\, at a typical radius of half the deconvolved $FWHM$ size of 0.025\,pc.  
We adopt thus $T_{\rm d}$=25\,K 
which results up to a factor of two uncertainty in the mass estimate.

%
   \begin{figure}[]
   \centering
   \includegraphics[width=3.7cm,angle=90]{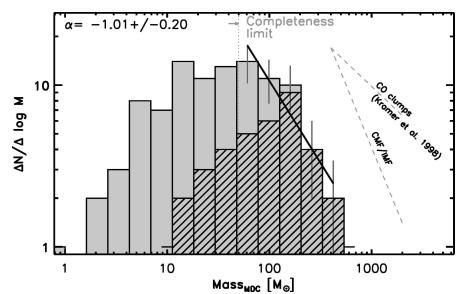}
      \caption{\small
Mass distribution of MDCs within $d$$\leq$4.5\,kpc. The Poisson error of each bin is shown as a grey line above the 10$\sigma_{\rm rms}$ completeness limit of 50\,\msol, the power-law fit is shown in a solid black line.
      Hashed area shows the distribution of the brightest cores ($M^{\rm max}_{\rm MDC}$) per clump. Dashed lines show the slope of the CMF/IMF \citep{Andre2014}, 
      and CO clumps \citep{Kramer1998}. }
         \label{fig:flux_distr}
   \end{figure}

The extracted cores have a mean mass of $\sim$63\,\msol\
corresponding to massive dense cores (MDCs as in \citealp{M07}),
and about 40\% of the sample hosts cores more massive than 150\,\msol. 
They are, in terms of physical properties, similar to SDC335-MM1 \citep{Peretto2013}, 
which is here the most massive core with 
$\sim$400\,\msol\ within a deconvolved FWHM size of $0.054$\,pc\footnote{
Our mass estimates for SDC335-MM1  
can be reconciled with \citet{Peretto2013} using a dust emissivity index of $\beta$$\sim$1.2 between 93\,GHz and 345\,GHz. A similarly low value of $\beta$  is also suggested by \citet{Avison2015}.}.
In these clumps the second brightest sources are also typically massive,  
on average 78\,\msol\ suggesting a preference to form more massive cores.
Except for one clump, no core is detected below 35\,\msol\
which is well above the typical detection threshold considering
the
mean 7$\sigma_{\rm rms}$ mass sensitivity of 11.2\,\msol\ at the mean distance of 2.6\,kpc, and may indicate a
lack of intermediate mass (between 10--40\,\msol) cores.
Similar findings have been reported towards a handful of other young massive sources by \citet{Bontemps2010} and \citet{Zhang2015}.  
Clumps with single sources host strictly massive cores with  $M_{\rm MDC}$$>$40\,\msol,
and about half of them reaches the highest mass range of $M_{\rm MDC}$$>$150\,\msol.

We show the mass distribution of cores as $\Delta N/\Delta {\rm log}\,M\sim M^{\alpha}$ in Fig.\,\ref{fig:flux_distr}, and indicate the 10$\sigma_{\rm rms}$ completeness limit of 50\,\msol, 
set by the highest noise in the poorest sensitivity data.
The distribution tends to be 
flat up to the completeness limit, and then shows a decrease at
the highest masses. 
The distribution of $M^{\rm max}_{\rm MDC}$  (hatched histogram) shows that the majority of the clumps host at least one massive core, 
while a few host only at most intermediate mass fragments. 
The least square power-law fit  
 to the highest mass bins above the completeness limit gives $\alpha=-1.01\pm0.20$, which is steeper than the distribution of CO clumps ($\alpha$=$-0.6$ to $-0.8$, \citealp{Kramer1998}), and tends to be shallower than the low-mass prestellar CMF and the stellar initial mass function (IMF) ($\alpha$=-1.35-- -1.5, \citealp{Andre2010}), although
at the high-mass end the scatter of the measured slopes is more significant \citep{Bastian2010}.
Using Monte Carlo methods we test the uncertainty of $\alpha$ due to the unknown
dust temperature, and simulated a range of $T_{\rm d}$ between 10$-$50\,K using a normal distribution with
 a mean of 25\,K,  and a power-law distribution. 
We fitted to the slope the same way, as above, and repeated the tests
until the standard deviation of the measured slope reached convergence. 
In good agreement with the observational results, { the
normal} temperature distribution gives $\alpha_{\rm MC}$=-1.01$\pm$0.11, and thus
constrains the error of the fit suggesting an intrinsically shallower slope than the IMF. 
A power-law temperature distribution in the same mass range with an exponent of $-0.5$,
could reproduce, however, the slope of the IMF, assuming that
the brightest sources are intrinsically warmer.
Alternatively, a larger level of fragmentation of the brightest cores
on smaller scales
could also reconcile our result with the IMF.

\section{Discussion}

%
   \begin{figure}[]
   \centering
   \includegraphics[width=5.0cm,angle=90]{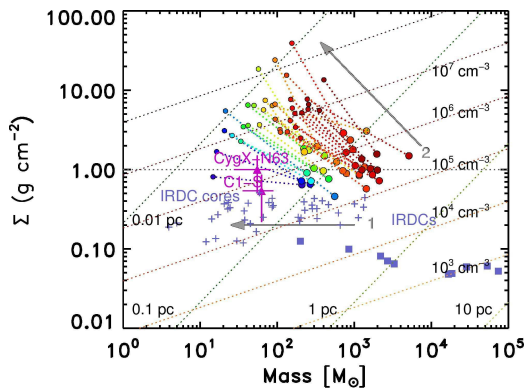}
      \caption{\small
      Surface-density versus mass diagram, coloured dotted lines in different shades show constant radius (green) and $n_{\rm H}$ number density (red) 
      (c.f.\,\citealp{Tan2014}).
Colored large circles show clumps (ATLASGAL), while smaller circles the cores (ALMA 7m array), colors scaling from blue to red with increasing $M^{\rm max}_{\rm MDC}$. We mark two massive cores with $M_{\rm MDC}=60$\,\msol\ (C1-S, \citealt{Tan2013}) and 55\,\msol\ (CygX-N63, \citealp{Bontemps2010}). For comparison IRDC clumps \citep{Kainulainen2013} and cores are shown \citep{BT2012}. Gray arrows show two models: 1) a uniform clump density, and 2)  a single central object with an $r^{-2}$ density profile.
              }
         \label{fig:surfdens}
   \end{figure}
   \begin{figure}[]
   \centering
   \includegraphics[width=5.0cm,angle=90]{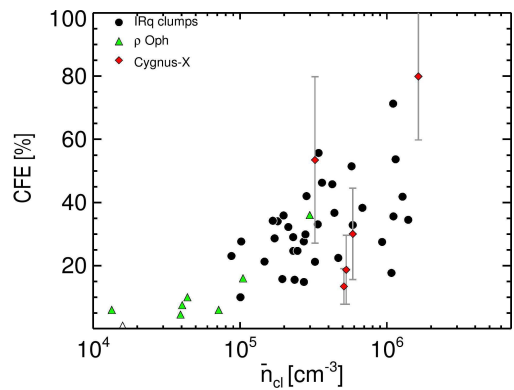}
         \caption{\small
      CFE versus average clump density ($\bar{n}_{\rm cl}$).
 Green triangles show cores of $\rho$~Oph \citep{M98} and red diamonds of Cygnus-X \citep{Bontemps2010}. 
              }
         \label{fig:cfe}
   \end{figure}

\subsection{Limited fragmentation from clump to core scale}\label{sec:monolothic_collapse}
 
The thermal Jeans mass in  massive clumps is low ($M_{\rm J}\sim$1\,\msol\ 
at $\bar{n}_{\rm cl}$=4.6$\times10^5$\,cm$^{-3}$, $T$=18\,K), which is expected to lead to
a high degree of fragmentation. 
In contrast, the observed 
infrared quiet massive clumps exhibit here limited fragmentation with $\bar{N}_{\rm fr}=3$,
from clump to core scales.
We even find single clumps/MDCs at our resolution. 
This is intriguing also because these most massive clumps of the Galaxy are expected to form rich clusters. 
The selected highest peak surface density clumps could therefore correspond to a phase of compactness  
where the large level of fragmentation to form a cluster has not yet developed. 

We find that the mass surface density ($\Sigma$) increases towards
small scales (Fig.\,\ref{fig:surfdens}, c.f.\,\citealp{Tan2014})
corresponding to a high concentration of mass.
80\% of the clumps host MDCs above 40\,\msol, and the
most massive fragments scale with the mass of their clump.
Two models are shown with arrows in Fig.\,\ref{fig:surfdens}:
1) clumps with a uniform mass distribution
forming low mass stars correspond to a roughly constant mass surface density;
2) clumps with all the mass concentrated in a single object corresponding to $n(r)\sim r^{-2}$ density profile. The majority of the sources fit better the steeper than uniform 
density profile.

The early fragmentation of massive clumps thus  
does not seem to follow thermal processes,
and shows fragment masses largely exceeding the local Jeans-mass 
 (see also \citealp{Zhang2009, Bontemps2010,Wang2014,Beuther2015, BT2012}). 
The significant concentration of mass on small scales also manifests in a high core formation efficiency (CFE), which { is the ratio of the total mass in fragments and the total clump mass from \citet{Csengeri2017} adopting the same physical parameters} (Fig.\,\ref{fig:cfe}). The CFE suggests an increasing concentration of mass in cores with the average clump volume density ($\bar{n}_{\rm cl}$), a trend which has been seen, although inferred from smaller scales, towards high-mass infrared quiet MDCs in Cygnus-X (\citealp{Bontemps2010}), and low-mass cores in $\rho$\,Oph (\citealp{M98}), and a sample of infrared bright MDCs (\citealp{Palau2013}). Although the CFE shows variations at high densities with $\bar{n}_{\rm cl}>10^5$\,cm$^{-3}$, exceptionally high CFE of over 50\%, can only be reached towards the highest average clump densities.

\subsection{Which physical processes influence fragmentation?}
\label{sec:stability}
 
What can explain that the thermal Jeans mass does not represent well the observed fragmentation properties in the early stages?
A combination of turbulence, magnetic field, and radiative feedback 
could increase the necessary mass scale for fragmentation.
Using the Turbulent Core model \citep{MT03}
for cores with $M_{\rm MDC}$$>$$150$\,\msol\ 
at the average radius of 0.025\,pc,
we estimate 
from their Eq.\,18 a turbulent line-width of $\Delta v_{\rm obs}$$\gtrsim$$6$\,\kms\ at the surface of cores,
which is a factor of two higher than the average $\Delta v_{\rm obs}$
at the clump scale \citep{Wienen2015}. The magnetic critical mass at the  
average clump density corresponds to $M_{\rm mag}$$<$400\,\msol\ at  
the typically observed magnetic field values of 1\,mG towards massive clumps
(e.g.\,\citealp{Falgarone2008,Girart2009,Cortes2016,Pillai2016})
following Eq.2.17 of \citet{BertoldiMcKee1992}. This suggests that
moderately strong magnetic fields could explain the large core masses, however,
at the high core densities of $\bar{n}_{\rm core}$=4$\times$10$^7$\,cm$^{-3}$
considerably stronger fields, at the order of B$>$10\,mG, would be required 
to keep the most massive cores subcritical. 
Although radiative feedback could also
limit fragmentation (e.g.\,\citealp{Krumholz2007, Longmore2011}), 
infrared quiet massive clumps are at the onset of star formation activity
and we lack evidence for a potential deeply embedded population of low-mass protostars
needed to heat up the collapsing gas. 

\subsection{Can global collapse explain the mass of MDCs?}

The rather monolithic fashion of collapse 
suggests that fragmentation is at least partly determined already at the clump scale,
which would be in 
agreement with observational signatures of global collapse of massive
filaments 
(e.g.\,\citealp{Schneider2010, Peretto2013}).
If entire cloud fragments undergo collapse, and equilibrium may not be reached on small scales
leading to the observed limited fragmentation and a high core formation efficiency at early stages. 
Mass replenishment beyond the clump scale could fuel the formation of the lower mass population of stars leading to an
increase in the number of fragments with time, and 
allowing a Jeans-like fragmentation to develop at more evolved stages (e.g.\,\citealp{Palau2015}). 

At the scale of cloud fragments, if collapse sets in at a lower density range of $\bar{n}_{\rm cloud}=10^2$\,cm$^{-3}$,
the initial thermal Jeans mass could reach $M_{\rm J}$$\sim$50\,\msol\ assuming T=18\,K, 
at a characteristic $\lambda_{\rm Jeans}$ of about 2.3\,pc.
This 
is consistent with the extent of globally collapsing clouds,
the involved mass range is, however, not sufficient to 
explain the mass reservoir of the most massive cores.
Considering the turbulent nature of molecular clouds 
in the form of large-scale flows, their shocks 
could compress larger extents of gas at higher
densities depending on the turbulent mach number (c.f.\,\citealp{Chabrier2011}),
and lead to an increase in the initial mass reservoir.
Fragmentation inhibition and the observed high CFE
are thus consistent with a collapse  setting in at  parsec scales.
The origin of their initial mass reservoir, however, still poses a challenge 
to current star formation models.  

\subsection{Towards the highest mass stars}
The mass distribution of MDCs could be reconciled with the IMF either if multiplicity prevailed on smaller than 0.06\,pc scales, or if the temperature distribution scales with the brightest fragments.  Similar results have been found towards
MDCs in Cygnus-X by \citet{Bontemps2010}, but also towards Galactic infrared-quiet clumps, such as G28.34+0.06 P1 \citep{Zhang2015}, and G11.11-0.12 P6 \citep{Wang2014}. 
Alternatively, the high core formation efficiency and a shallow core mass distribution could suggest 
an intrinsically top-heavy distribution of high-mass protostars at the early phases. 
Considering the twelve highest mass cores with $M_{\rm MDC}$=150$-$400\,\msol\ and an efficiency ($\epsilon$) of $10-30$\% (e.g.\,\citealp{Tanaka2016}), we could expect a population of stars with a final stellar mass of $M_{\star}\sim \epsilon\times M_{\rm MDC}=15-120$\,\msol, reaching the highest mass O-type stars.

\section{Conclusions}
We study the fragmentation of a representative selection of a homogenous sample of massive
infrared-quiet clumps, and reveal a population of MDCs reaching up to $\sim$400\,\msol.
A large fraction (77\%) of clumps exhibit limited fragmentation, 
and host MDCs. 
The fragmentation of massive clumps  
 suggests a large concentration of mass at small scales and a high CFE. 
We lack observational support for strong enough turbulence and magnetic field to keep
the most massive cores virialized.  
Our results are consistent with entire cloud fragments in global collapse, 
while the origin of their pre-collapse mass reservoir still challenges current star formation models.

\begin{acknowledgements}
We thank the referee for constructive comments on the manuscript. This paper makes use of the ALMA data: ADS/JAO.ALMA 2013.1.00960.S. ALMA is a partnership of ESO (representing its member states), NSF (USA) and NINS (Japan), together with NRC (Canada), NSC and ASIAA (Taiwan), and KASI (Republic of Korea), in cooperation with the Republic of Chile. The Joint ALMA Observatory is operated by ESO, AUI/NRAO and NAOJ. T.Cs. acknowledges support from the \emph{Deut\-sche For\-schungs\-ge\-mein\-schaft, DFG\/}  via the SPP (priority programme) 1573 'Physics of the ISM'. 
HB acknowledges support from the European Research Council under the Horizon 2020 framework program via the ERC Consolidator Grant CSF-648505. LB acknowledges support from CONICYT PFB-06 project. 
 A.P. acknowledges financial support from UNAM-DGAPA-PAPIIT IA102815 grant, M\'exico.
 \end{acknowledgements}

\vspace{-1cm}
   \bibliographystyle{aa} 
   \bibliography{aca-letter.bib} 

\begin{appendix} 
   \begin{figure*}[!htpb]
   \centering
   \includegraphics[width=0.25\linewidth]{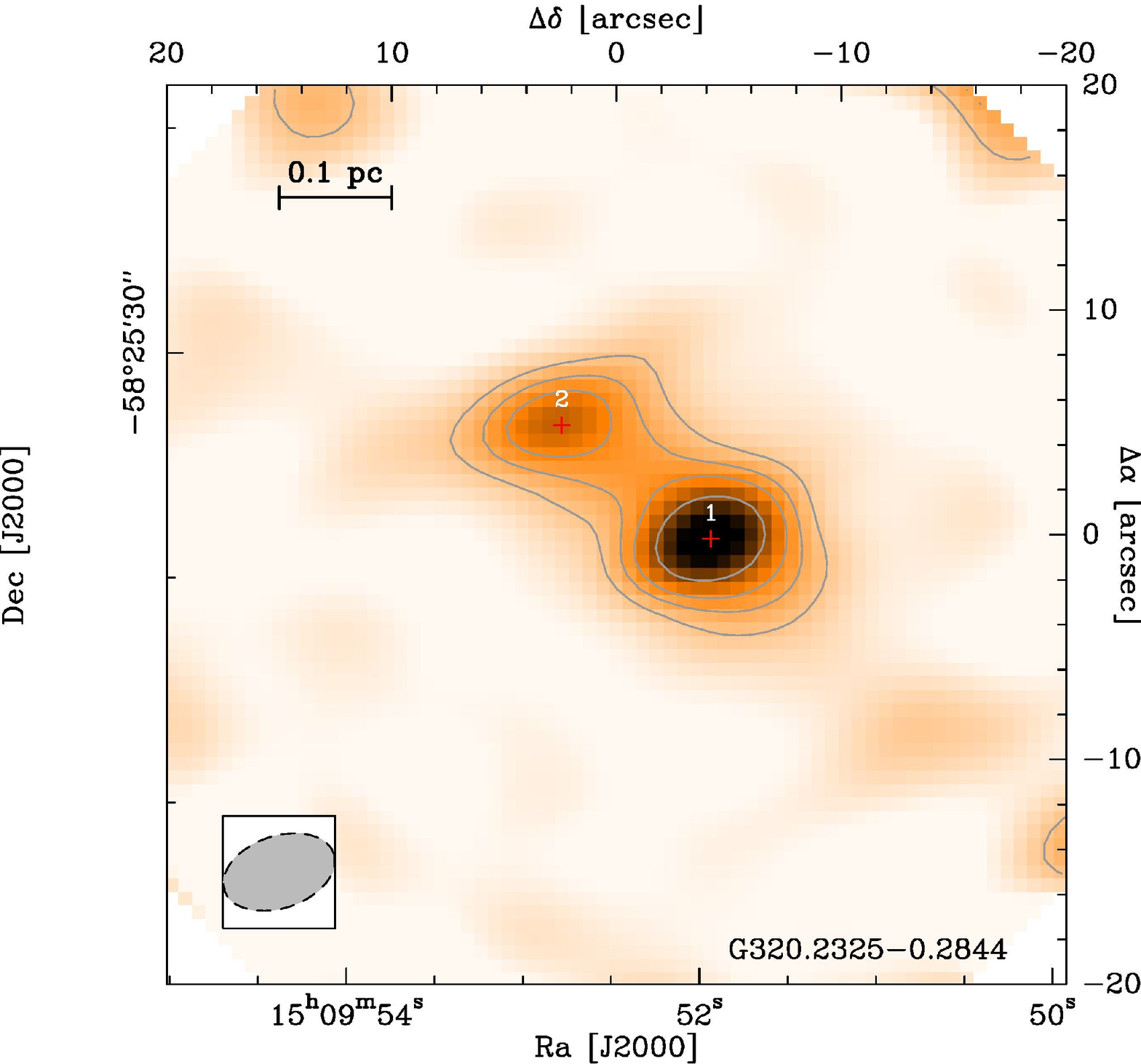}
   \includegraphics[width=0.25\linewidth]{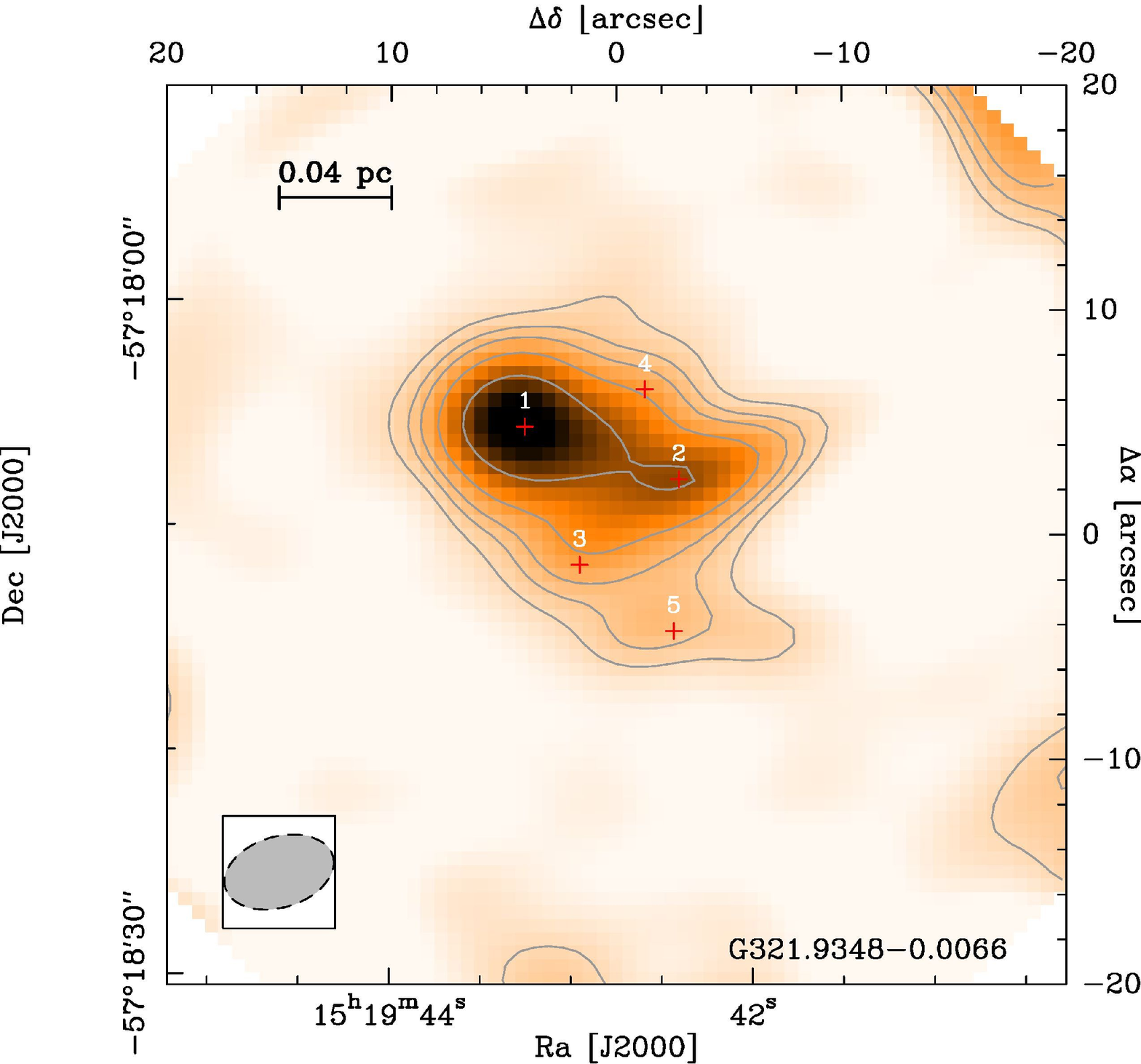}
   \includegraphics[width=0.25\linewidth]{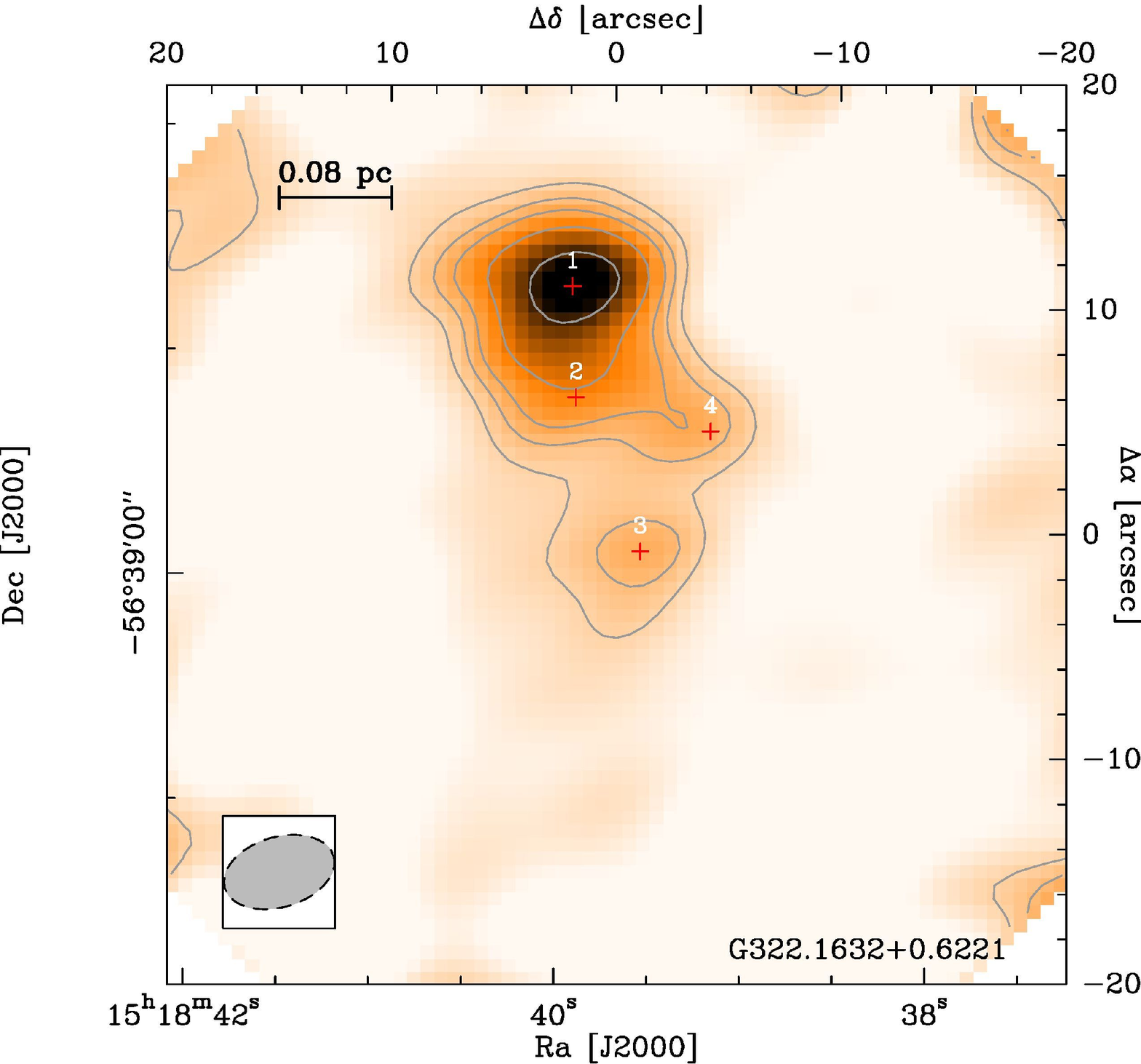}
   \includegraphics[width=0.25\linewidth]{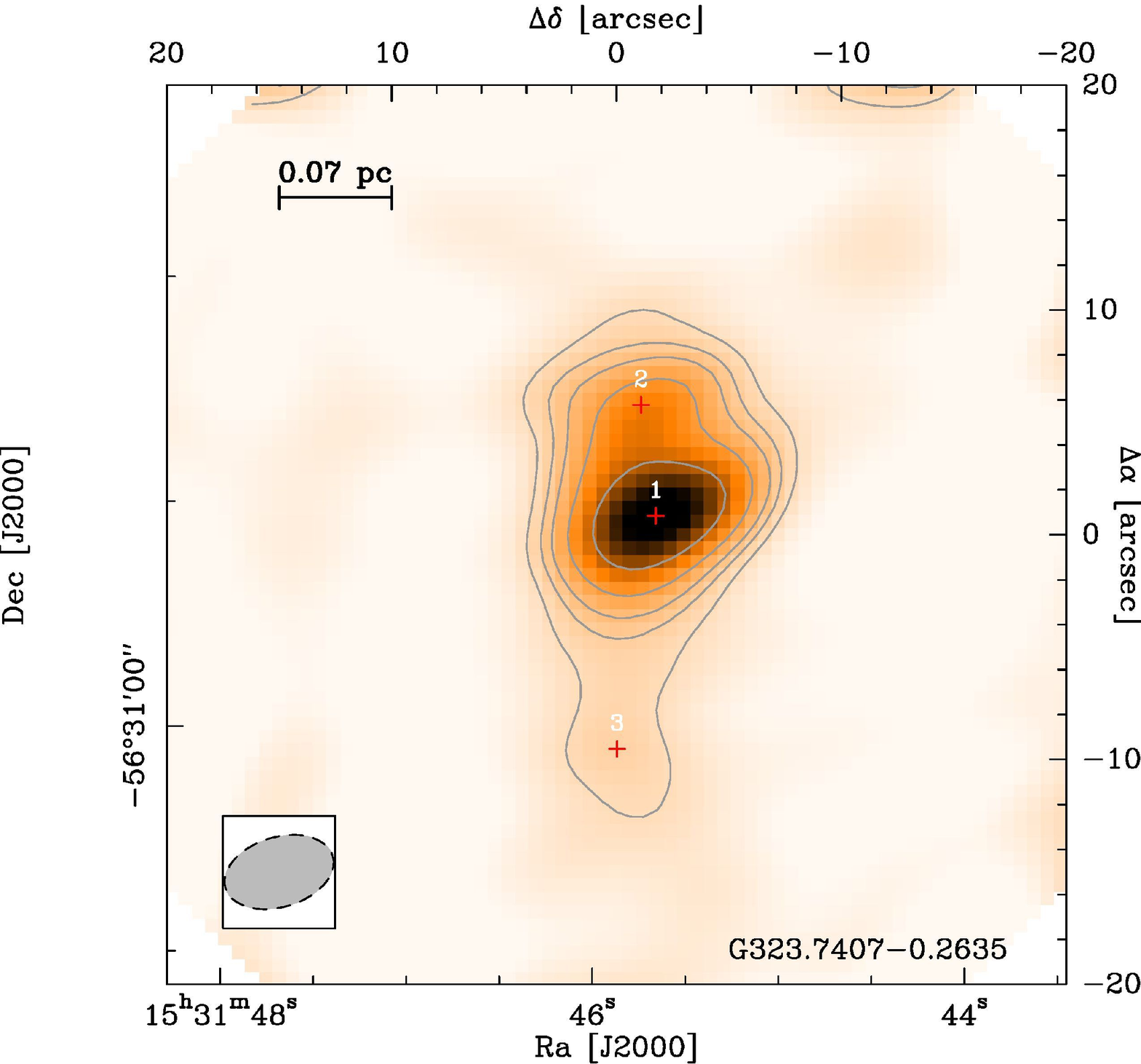}
   \includegraphics[width=0.25\linewidth]{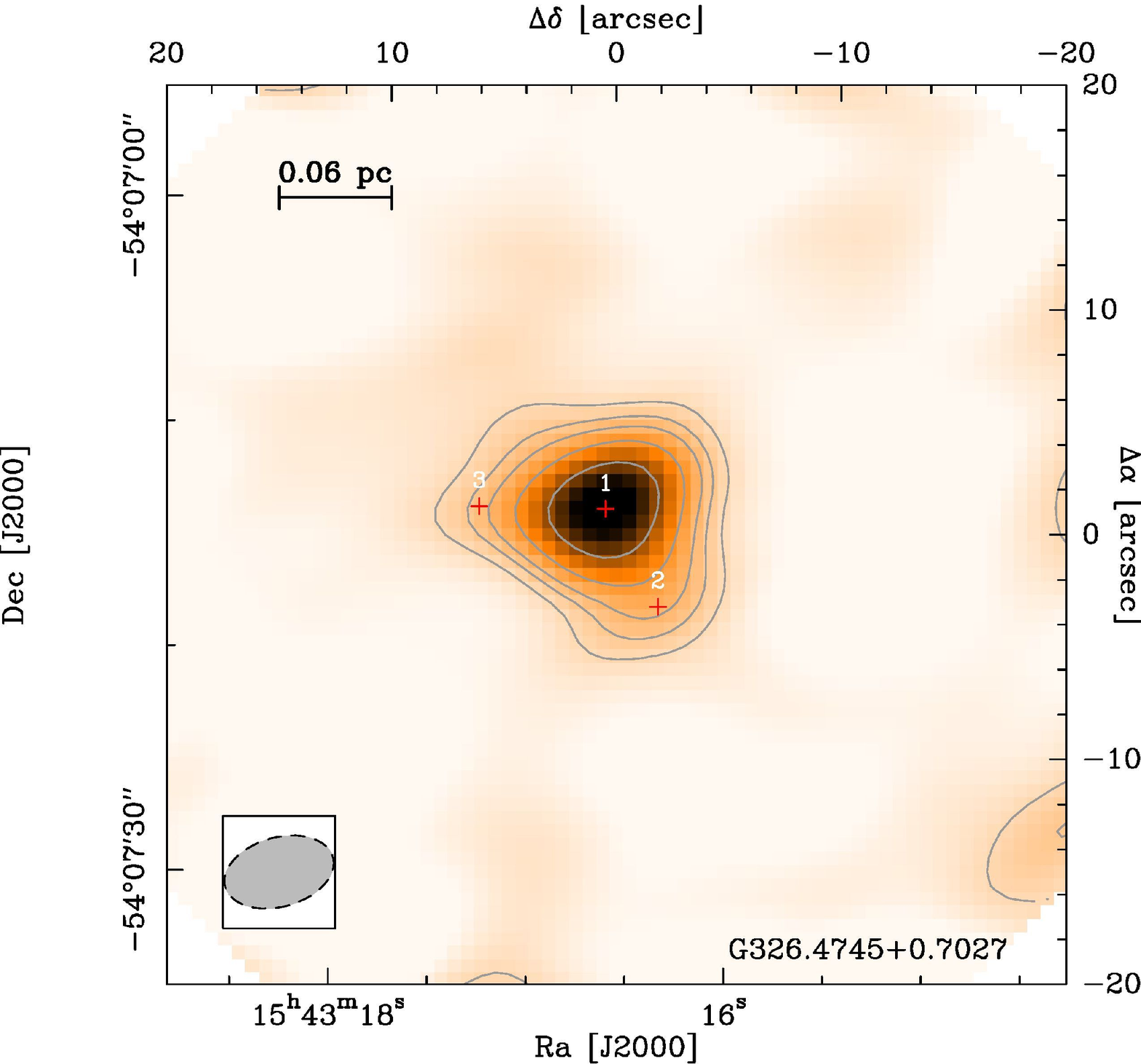}
   \includegraphics[width=0.25\linewidth]{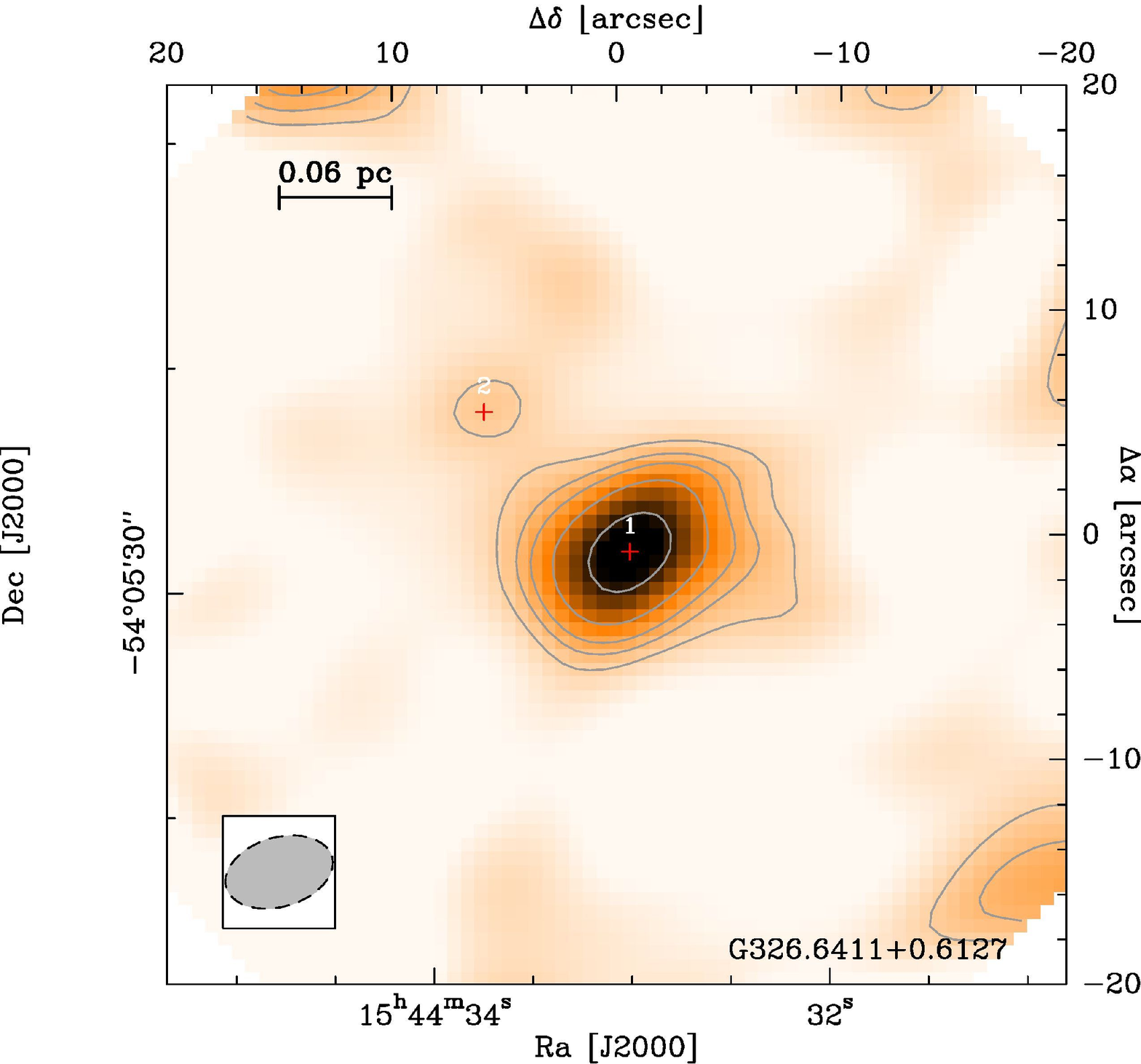}
   \includegraphics[width=0.25\linewidth]{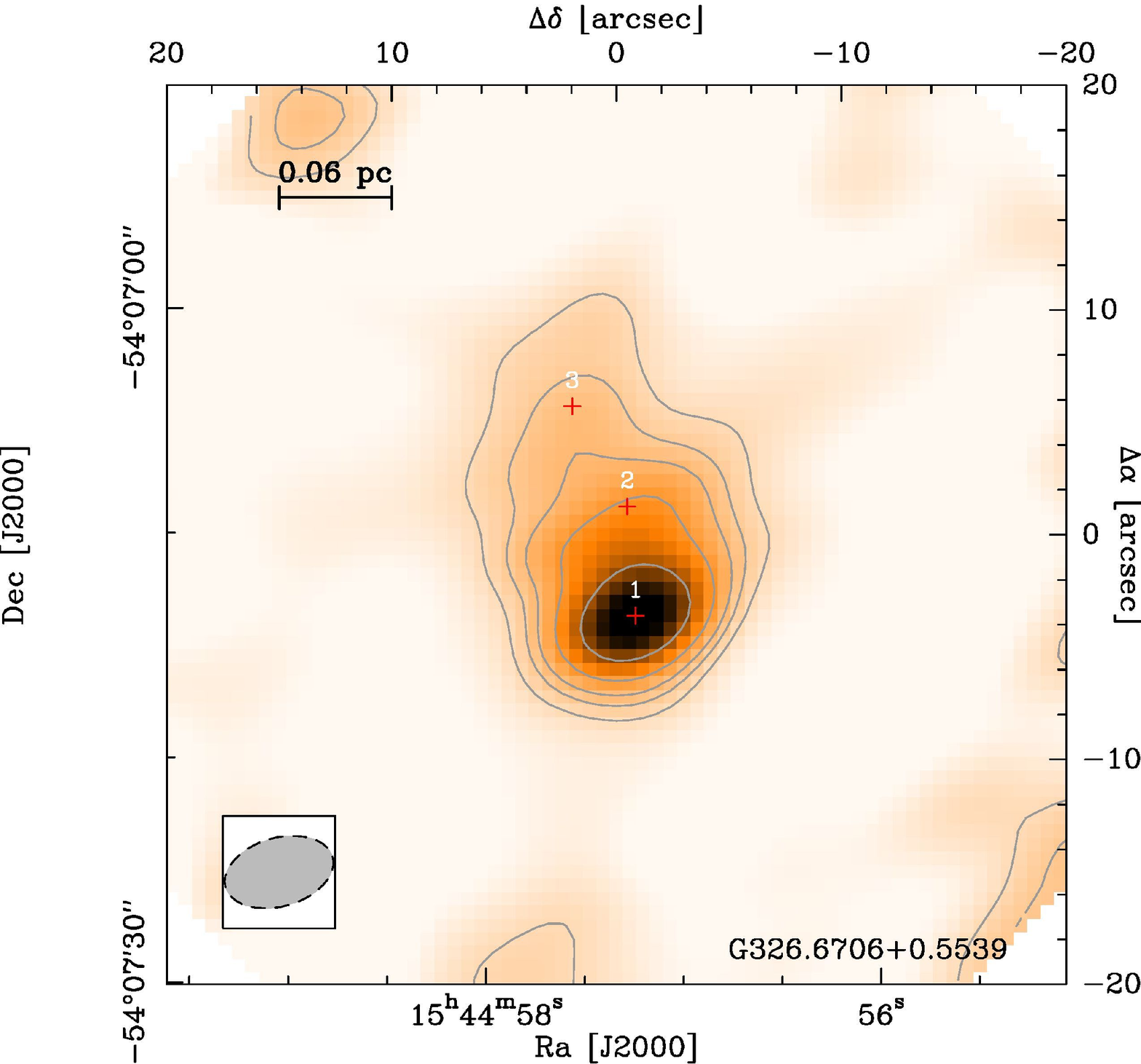}
   \includegraphics[width=0.25\linewidth]{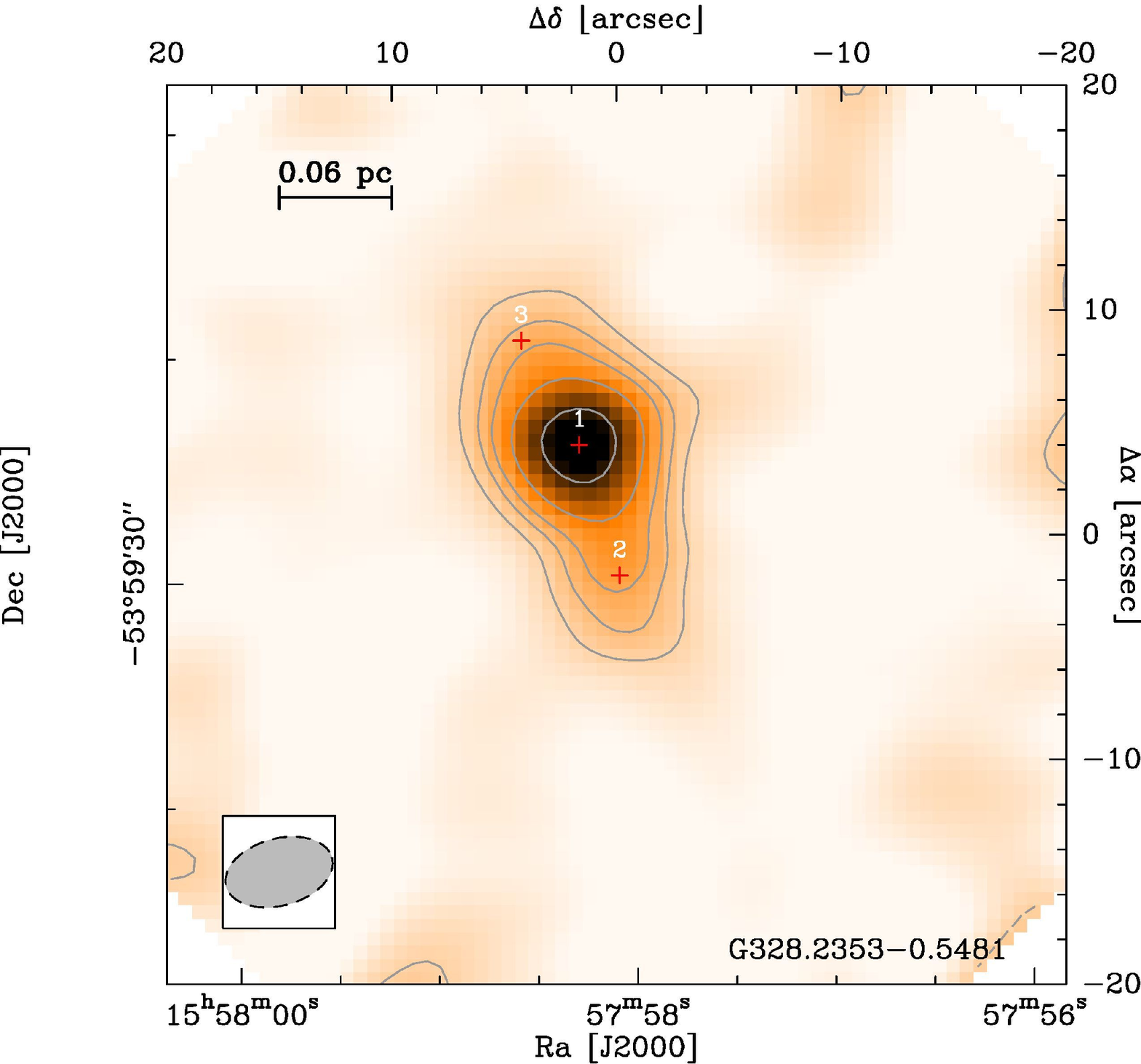}
   \includegraphics[width=0.25\linewidth]{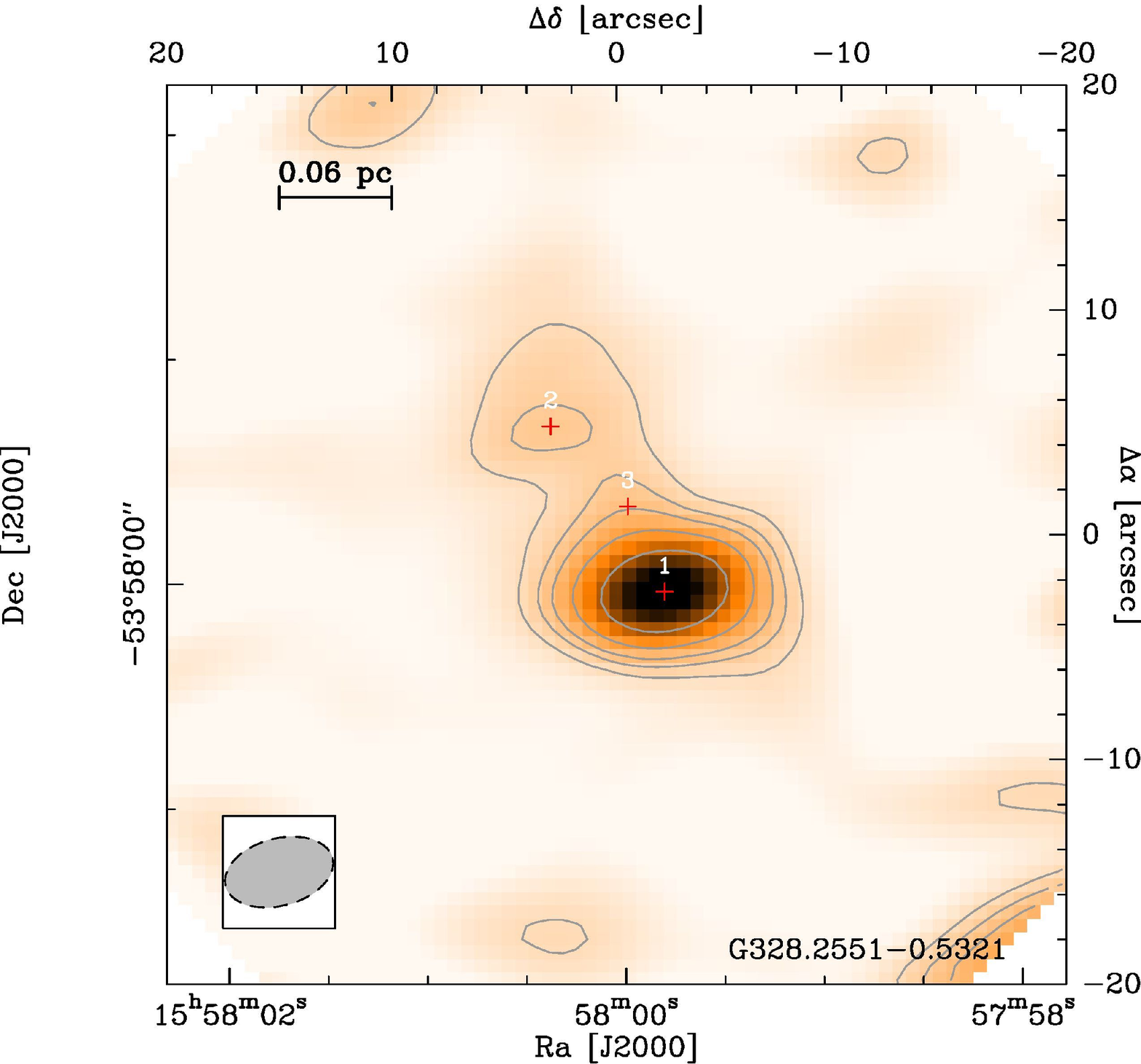}
   \includegraphics[width=0.25\linewidth]{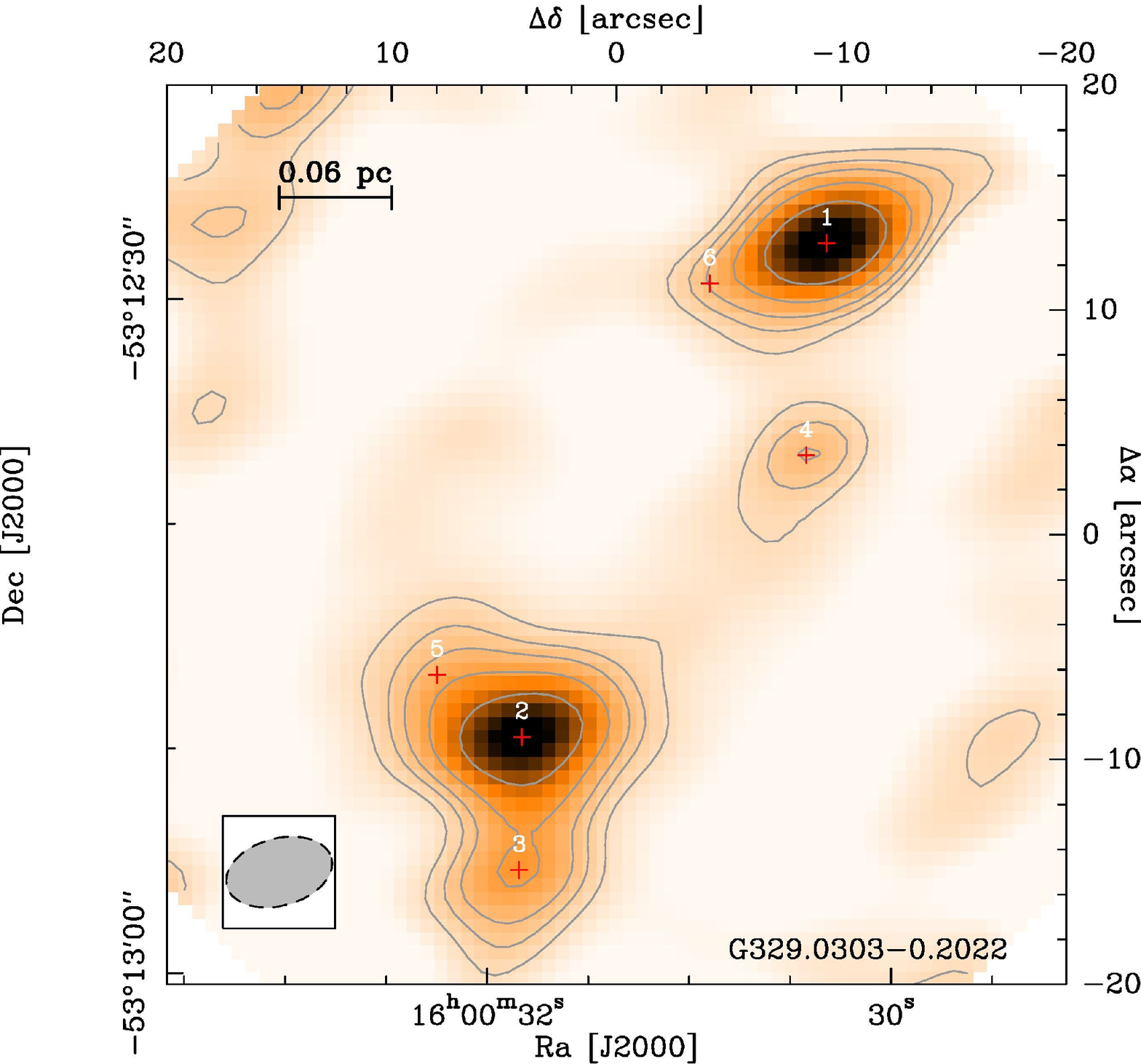}
   \includegraphics[width=0.25\linewidth]{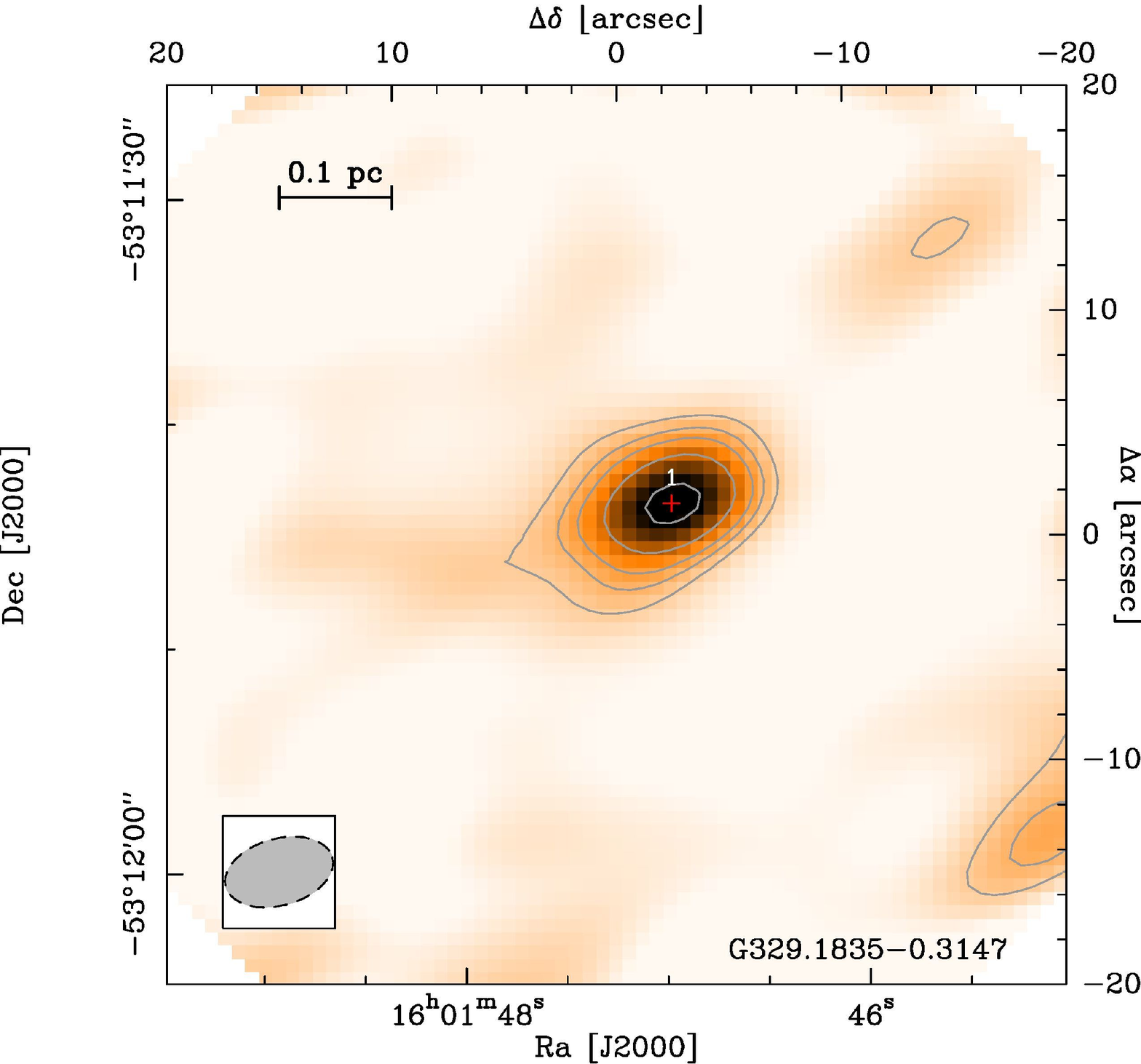}
    \includegraphics[width=0.25\linewidth]{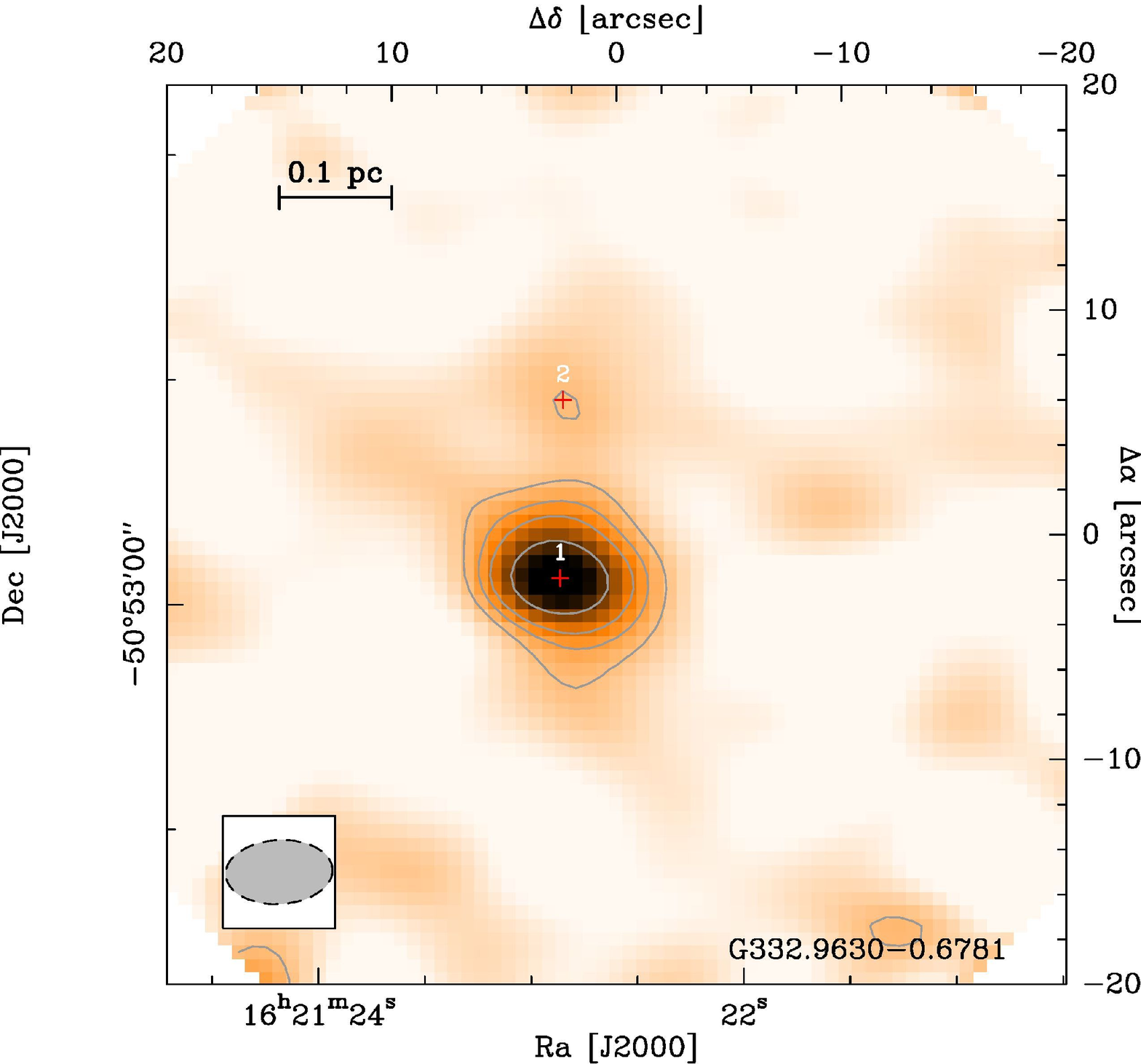}
    \includegraphics[width=0.25\linewidth]{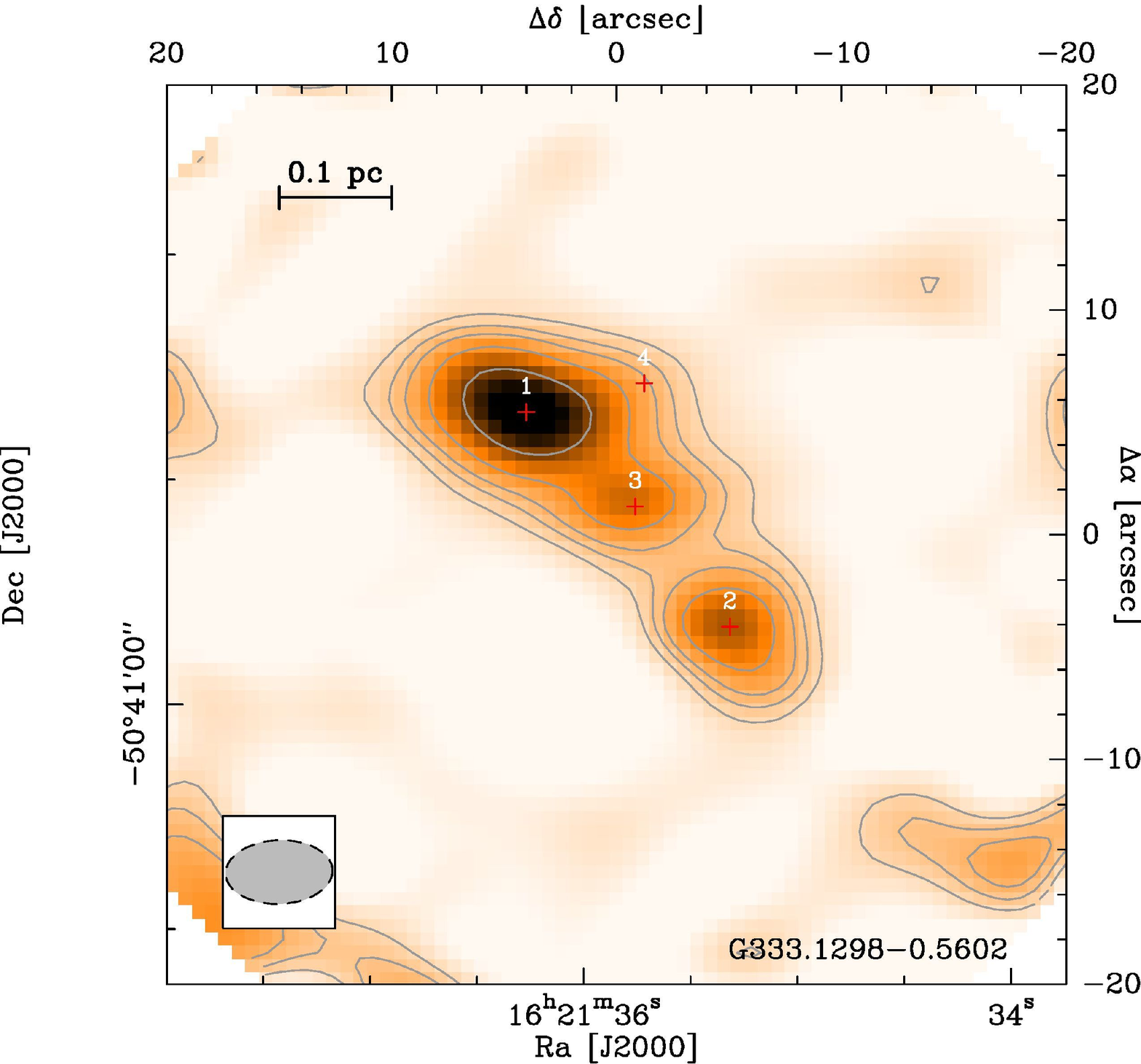}
    \includegraphics[width=0.25\linewidth]{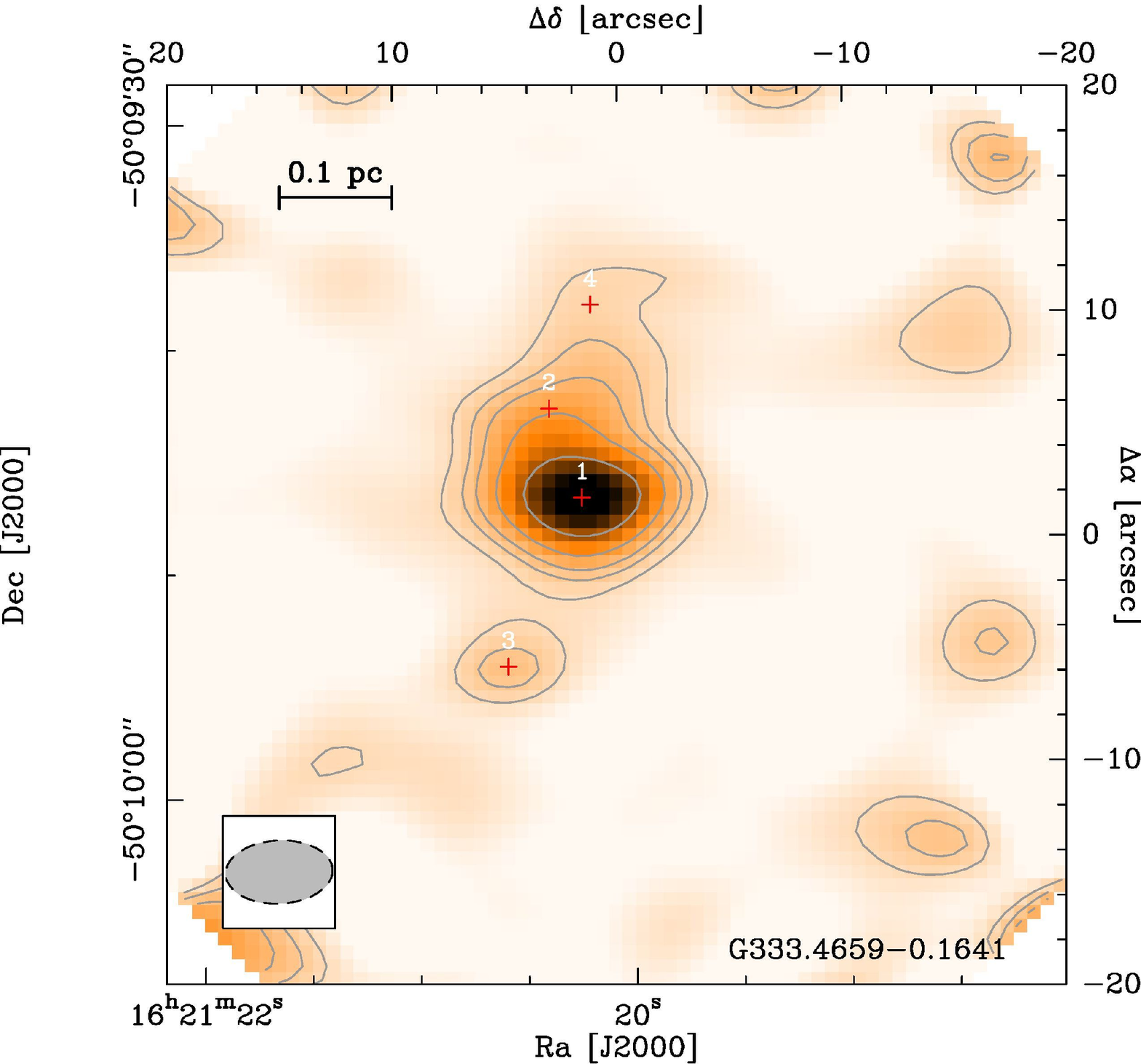}
    \includegraphics[width=0.25\linewidth]{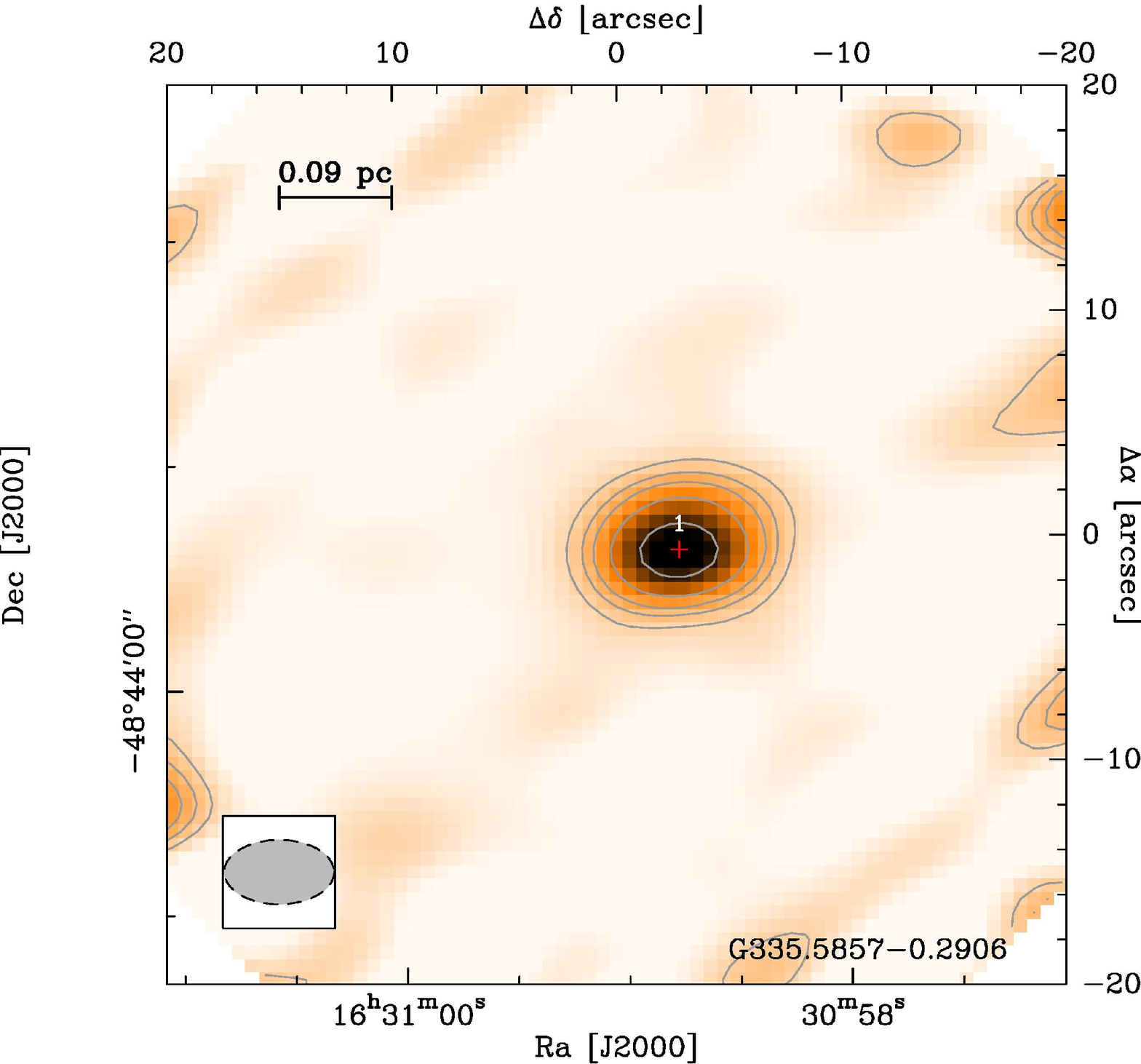}
       \caption{
   		Line-free continuum emission at 345\,GHz with ALMA 7m array. Contours start at $7\times$ the $rms$ noise and increase in a logarithmic scale. Red crosses mark the continuum sources with labels in white (see Table\,\ref{table:long}). The beam is shown in the lower left corner of each panel.}\label{fig:all}
    \end{figure*}

   \begin{figure*}[!htpb]
   \centering
   \ContinuedFloat
    \includegraphics[width=0.25\linewidth]{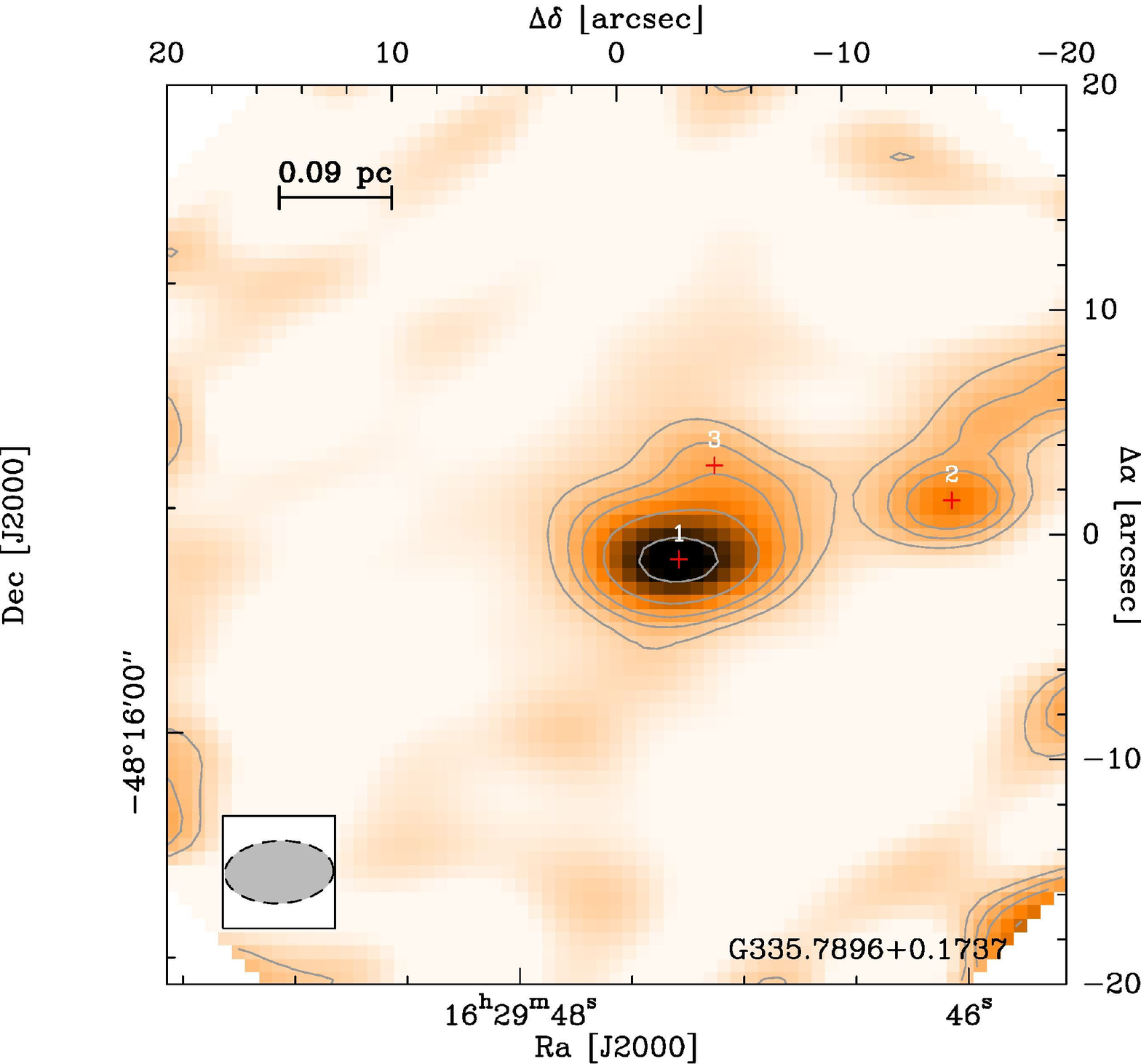}
   \includegraphics[width=0.25\linewidth]{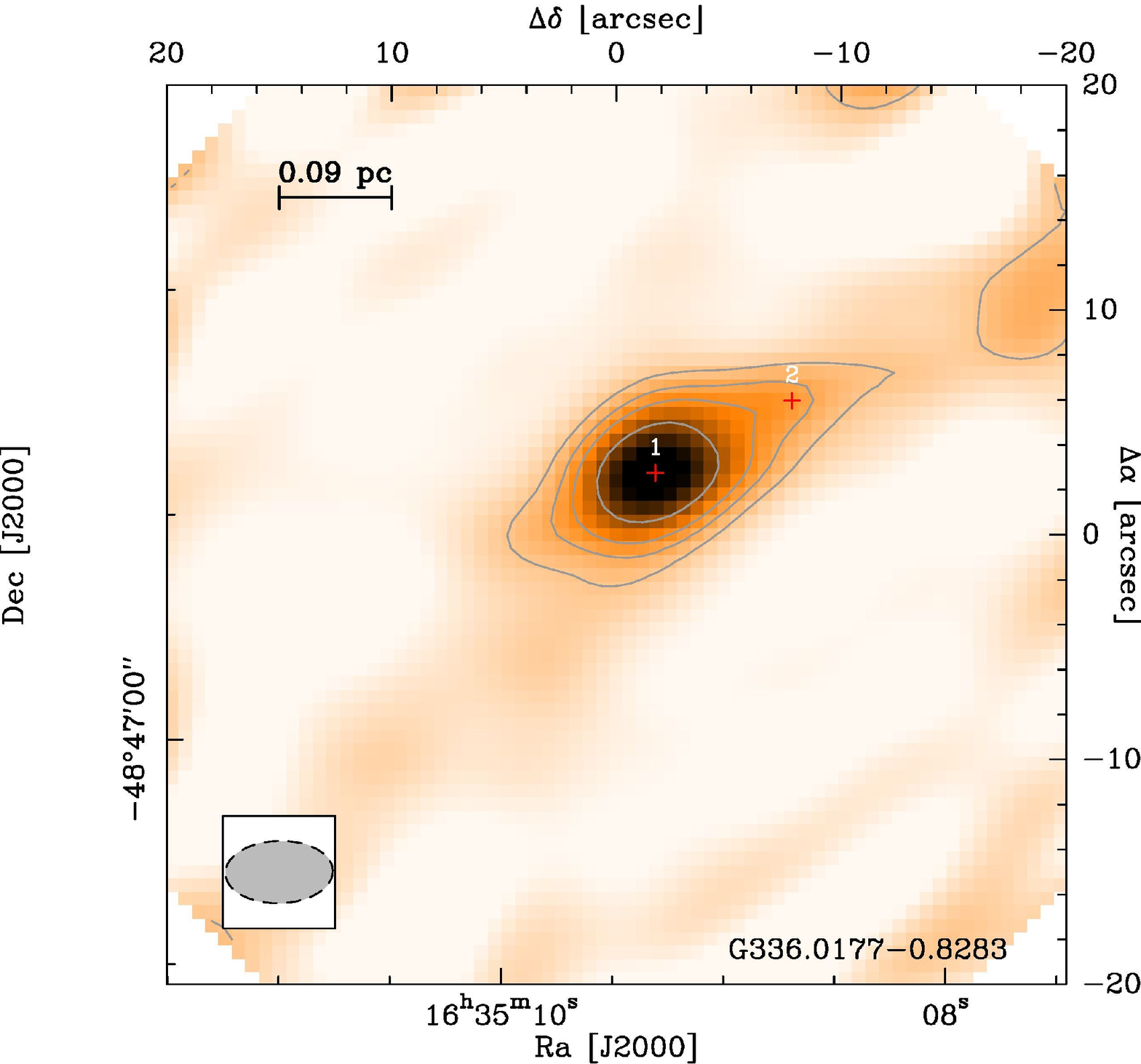}
     \includegraphics[width=0.25\linewidth]{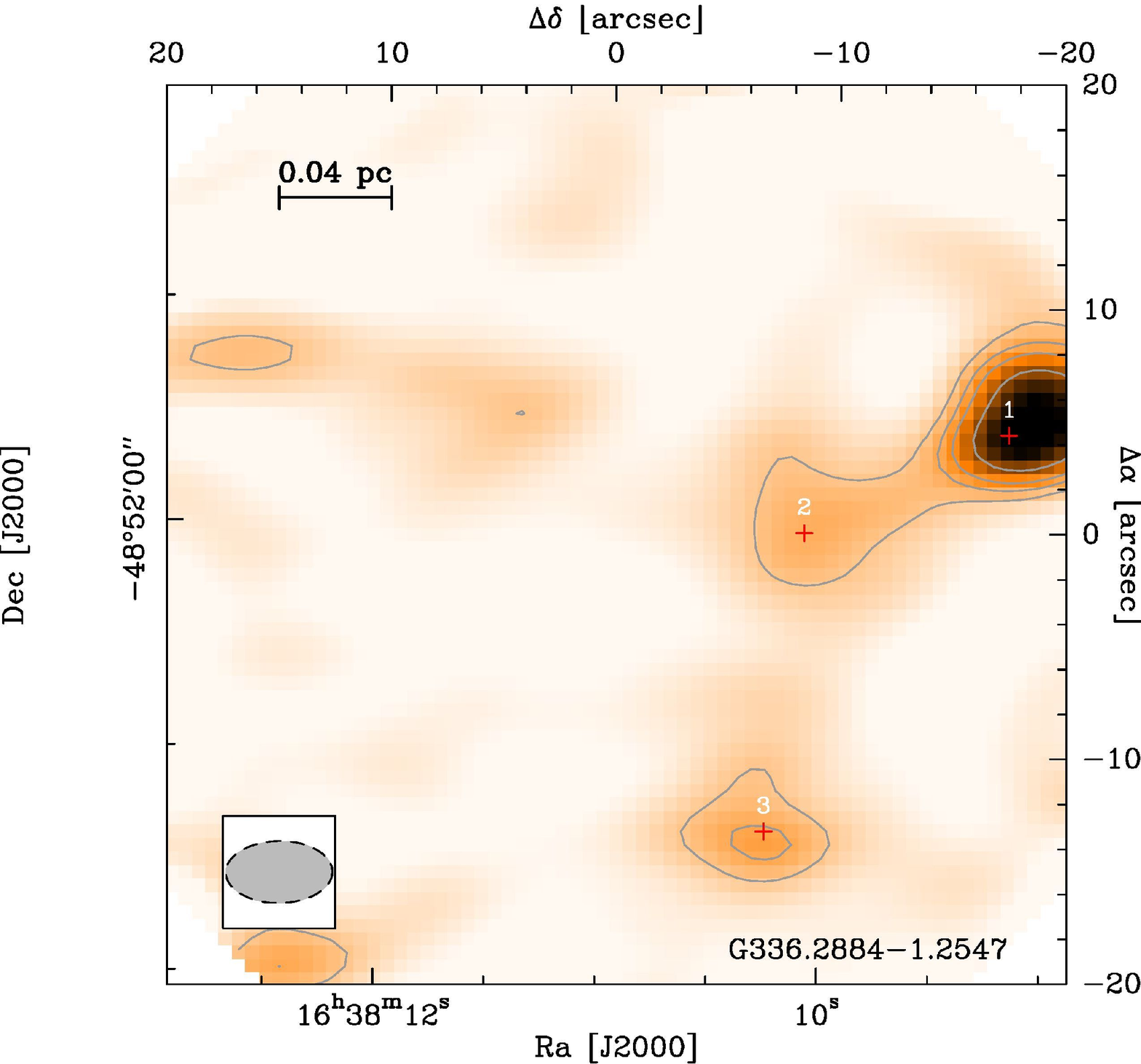}
    \includegraphics[width=0.25\linewidth]{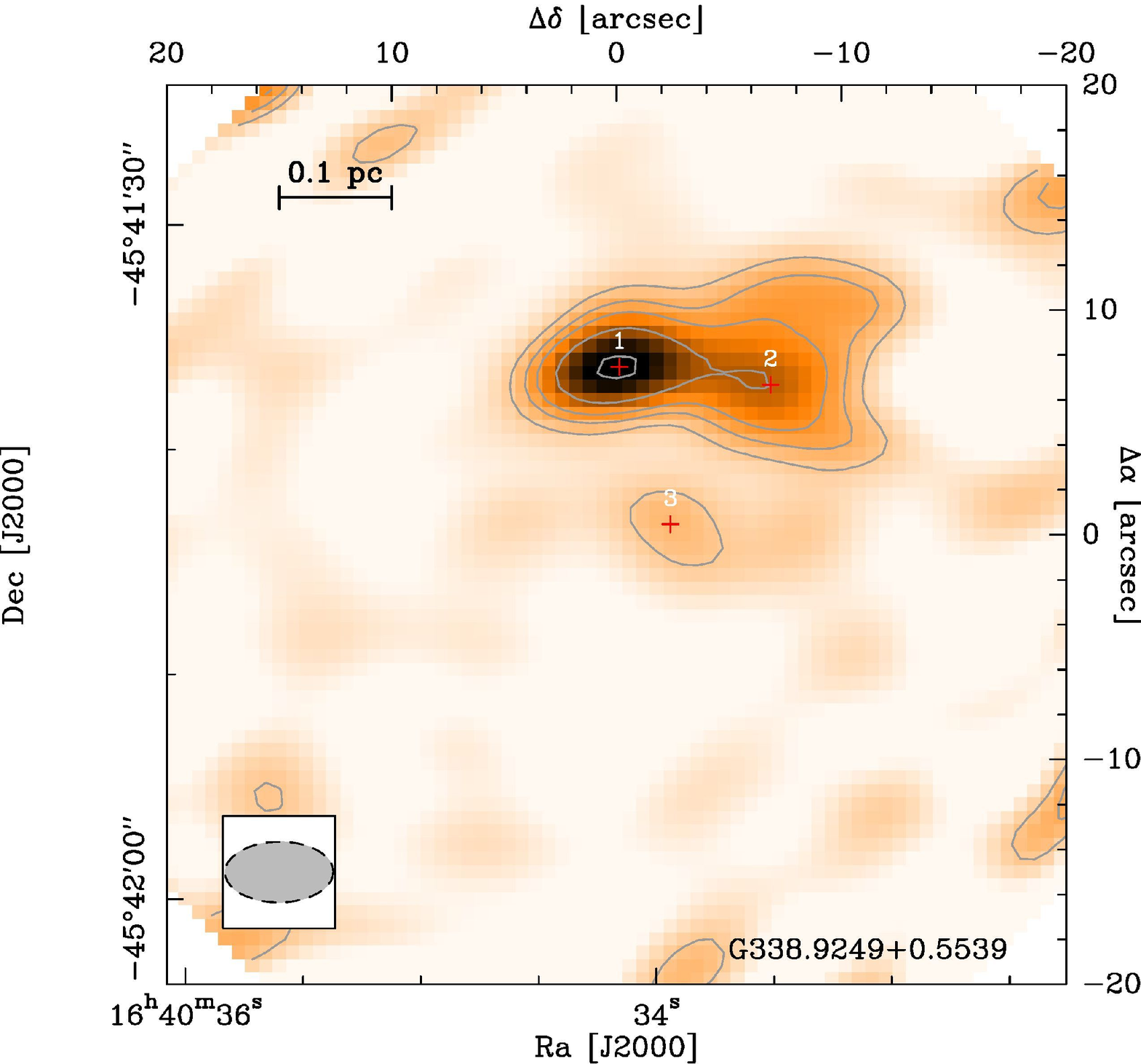}
    \includegraphics[width=0.25\linewidth]{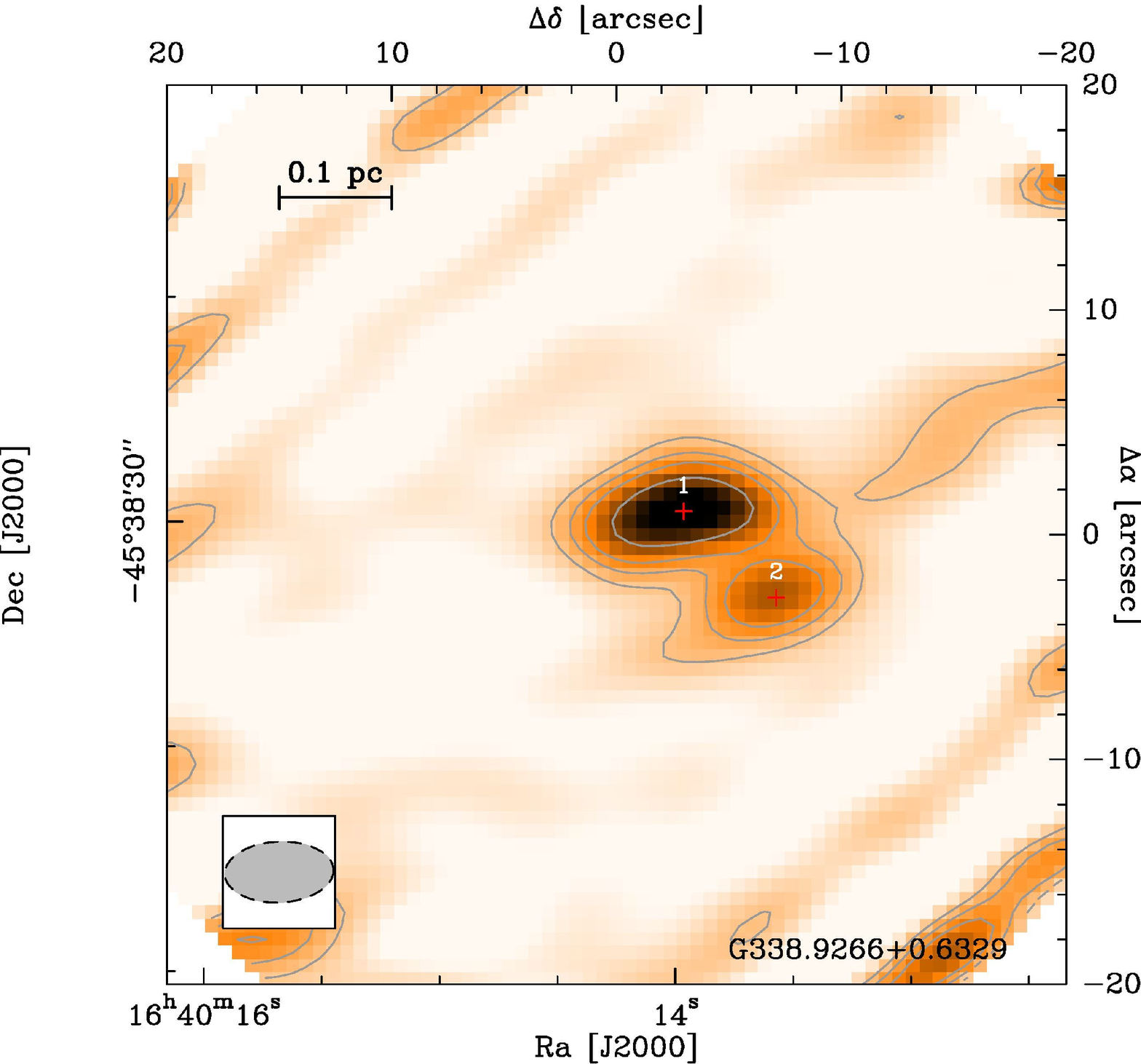}
    \includegraphics[width=0.25\linewidth]{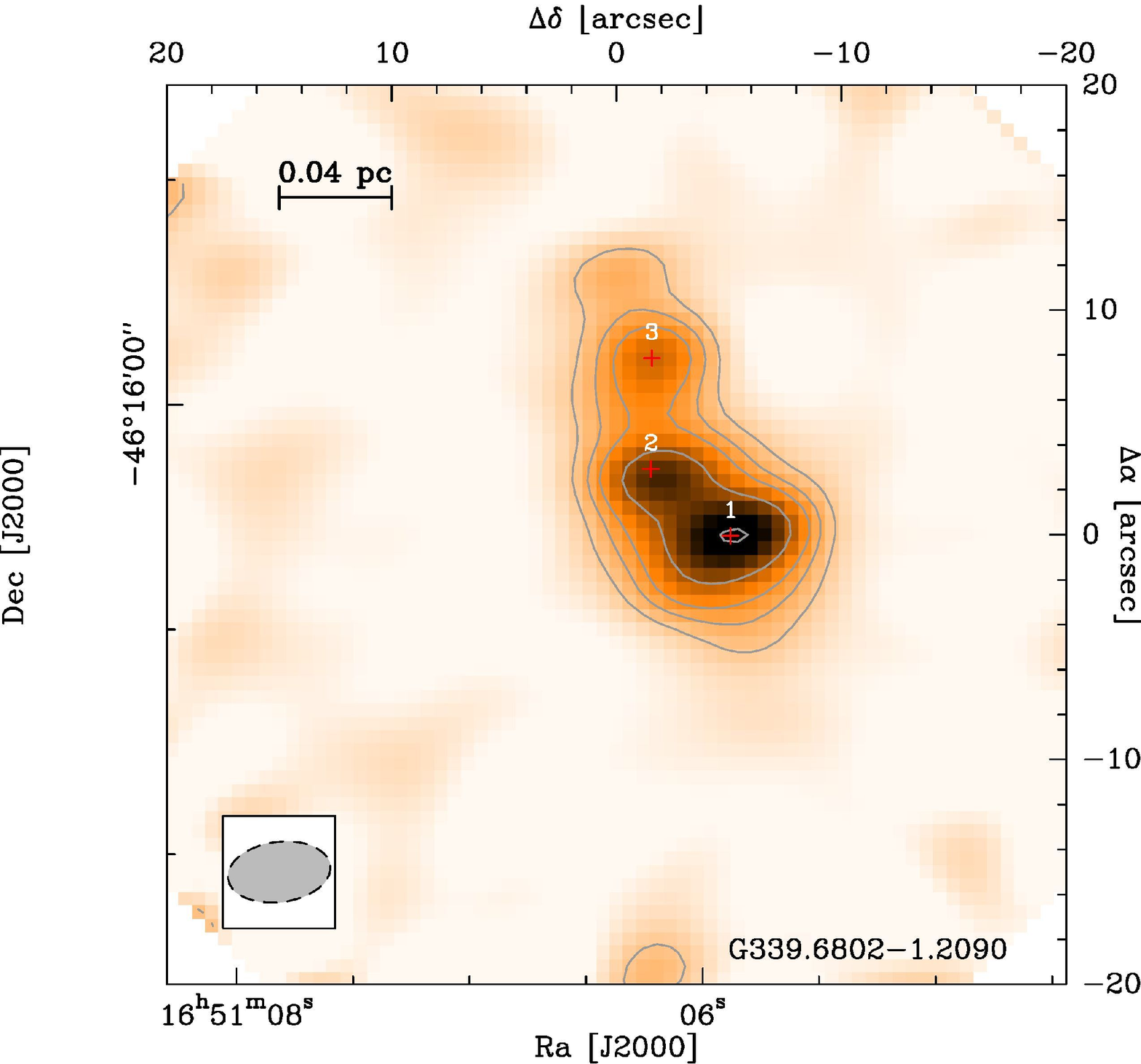}
    \includegraphics[width=0.25\linewidth]{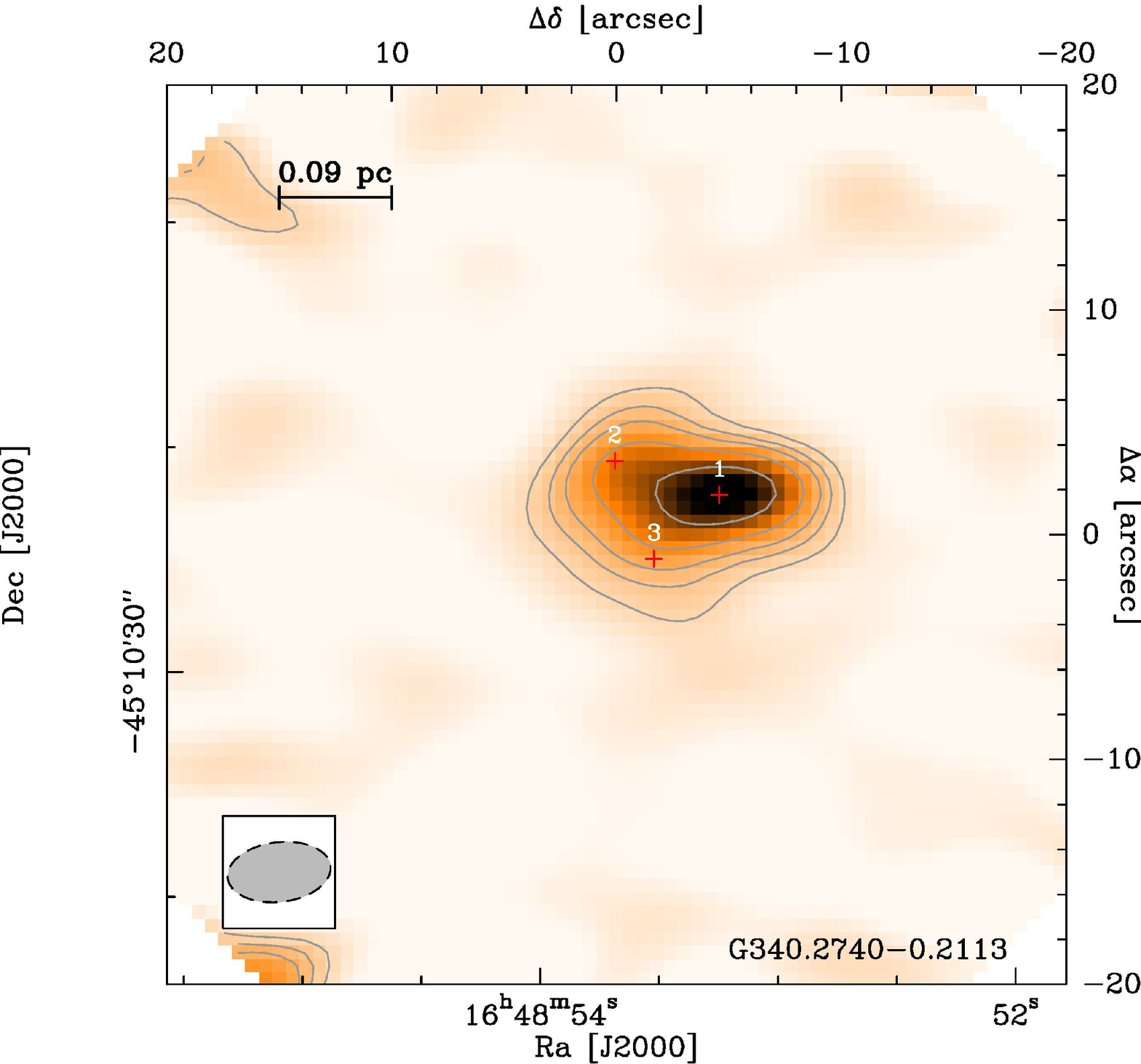}
    \includegraphics[width=0.25\linewidth]{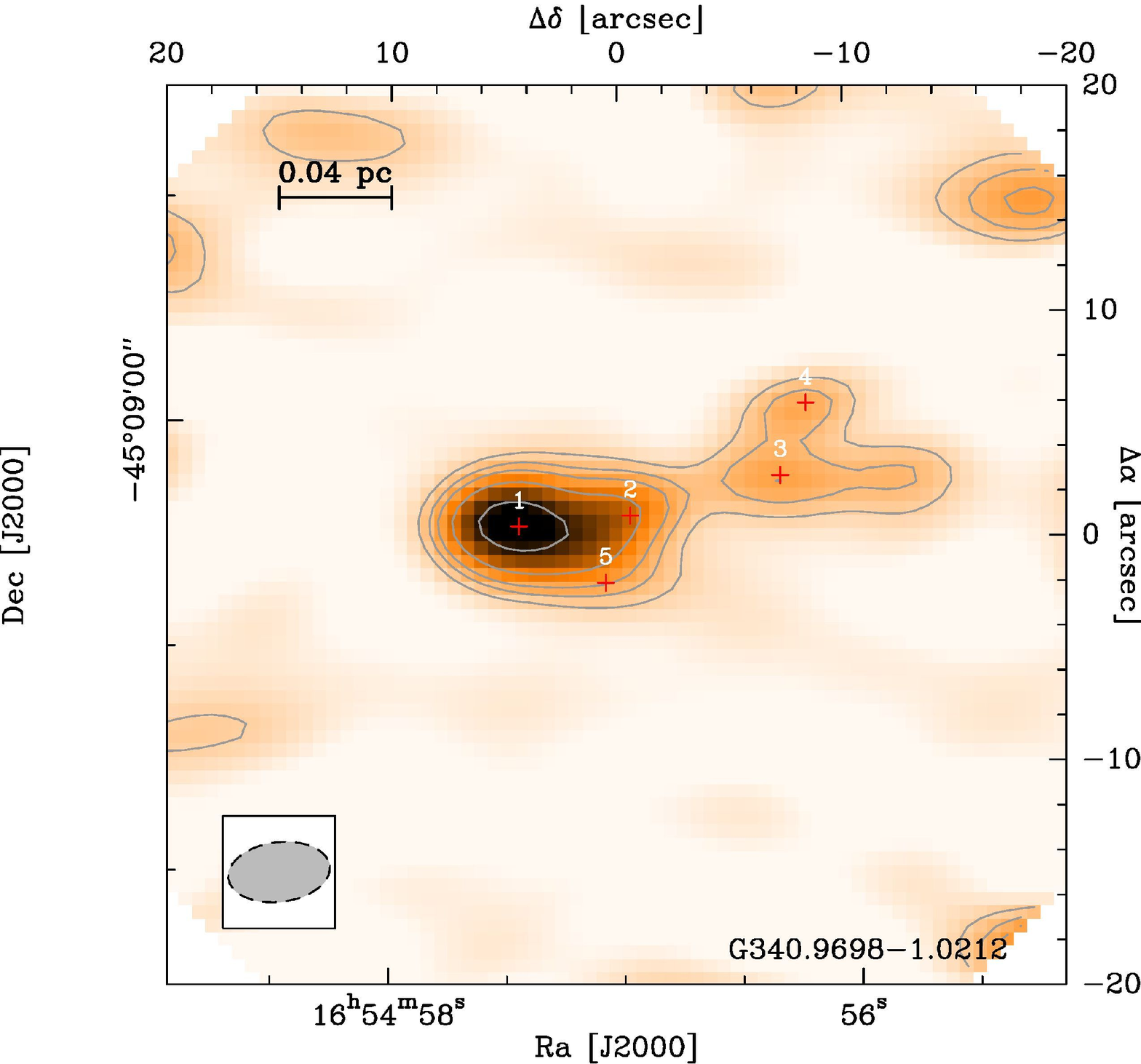}
    \includegraphics[width=0.25\linewidth]{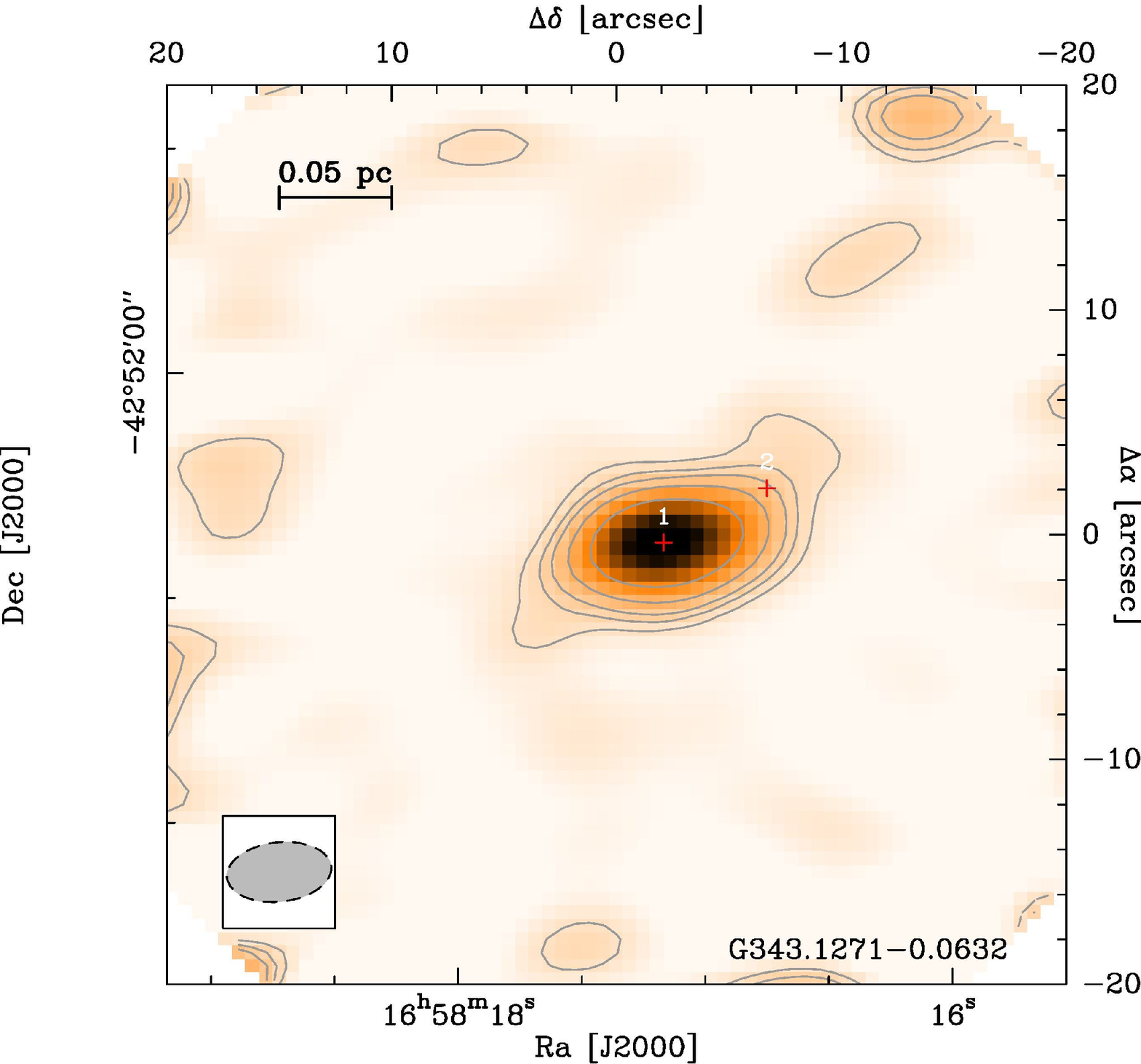}
    \includegraphics[width=0.25\linewidth]{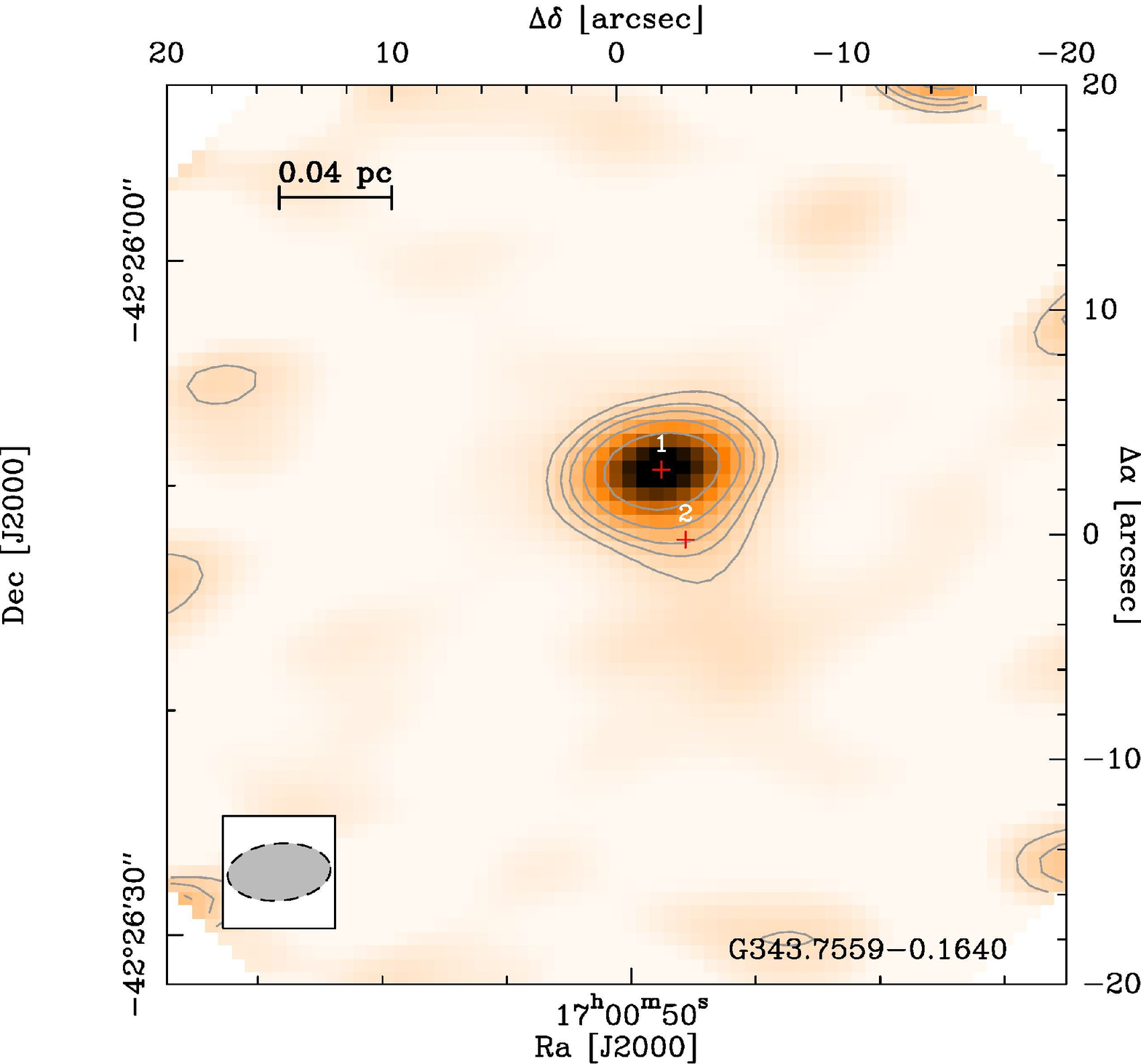}
    \includegraphics[width=0.25\linewidth]{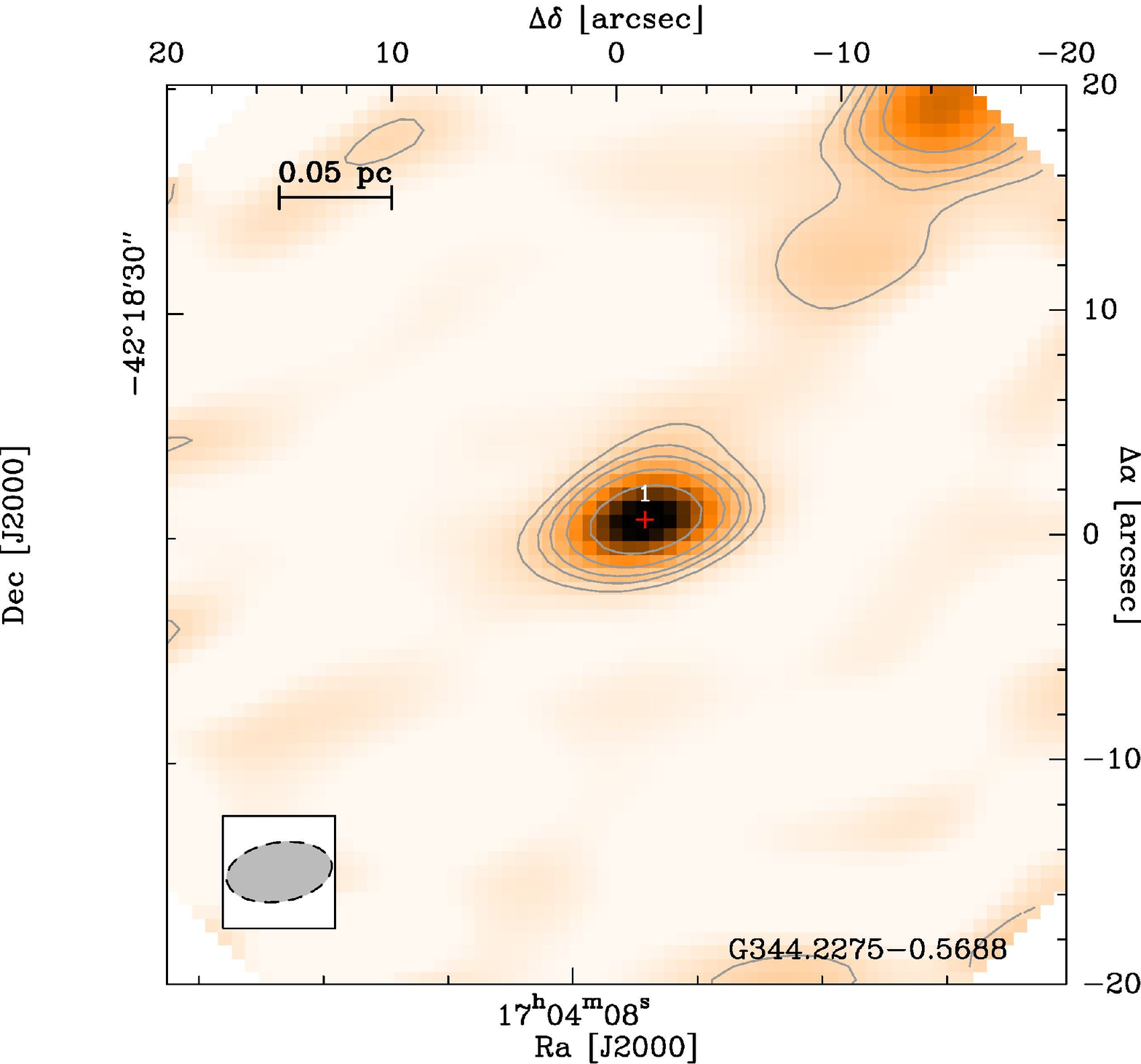}
    \includegraphics[width=0.25\linewidth]{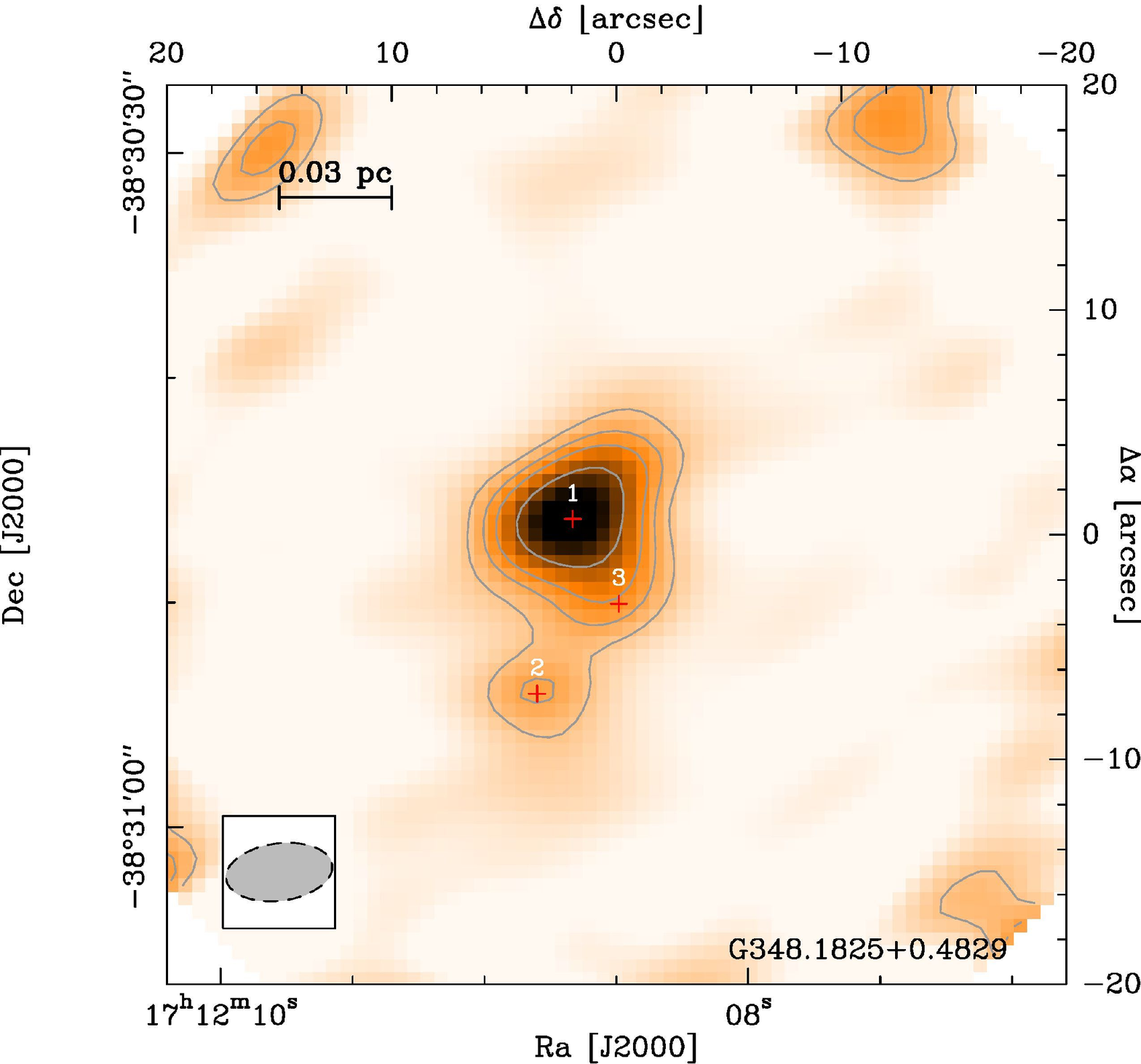}
    \includegraphics[width=0.25\linewidth]{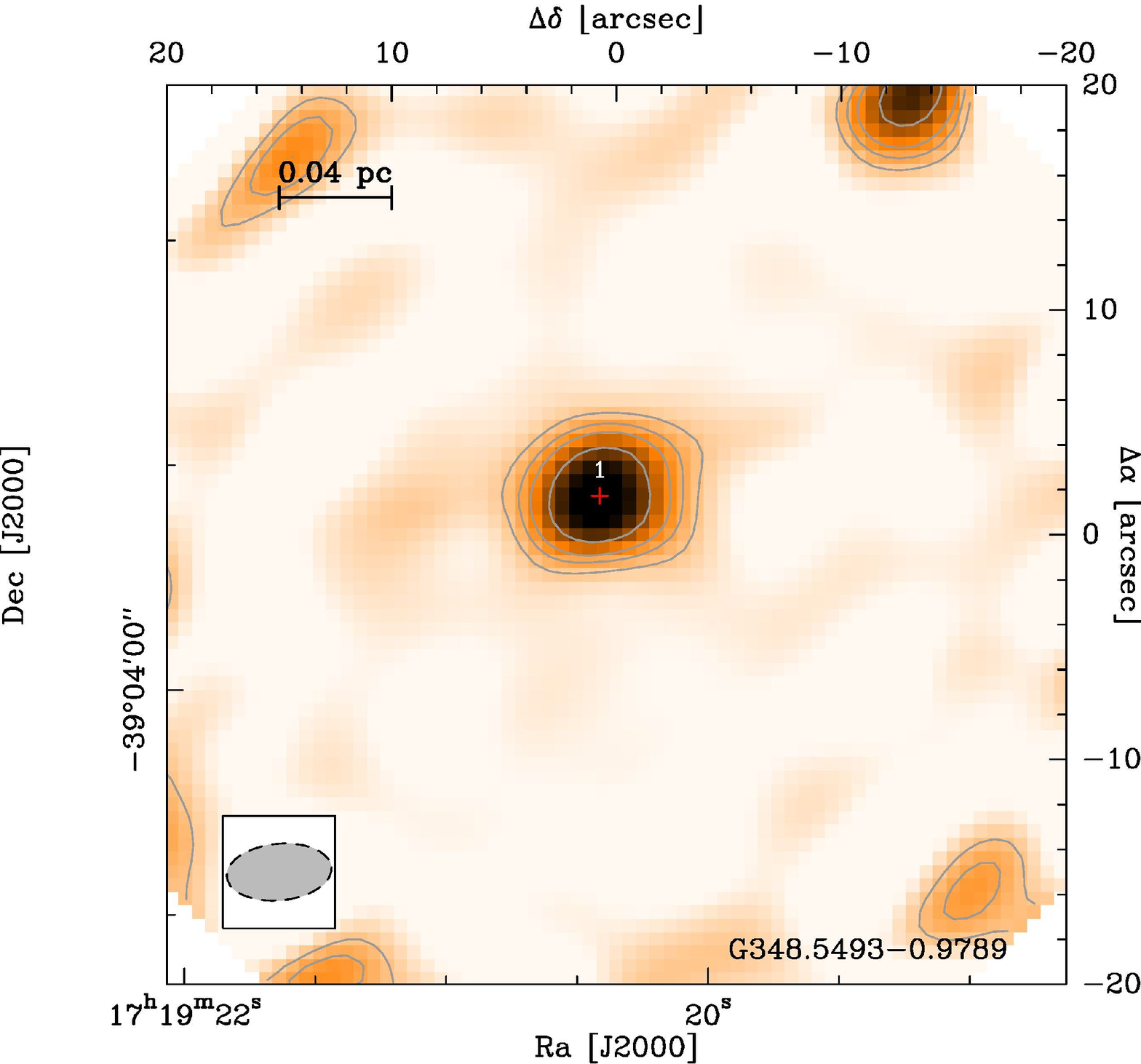}
    \includegraphics[width=0.25\linewidth]{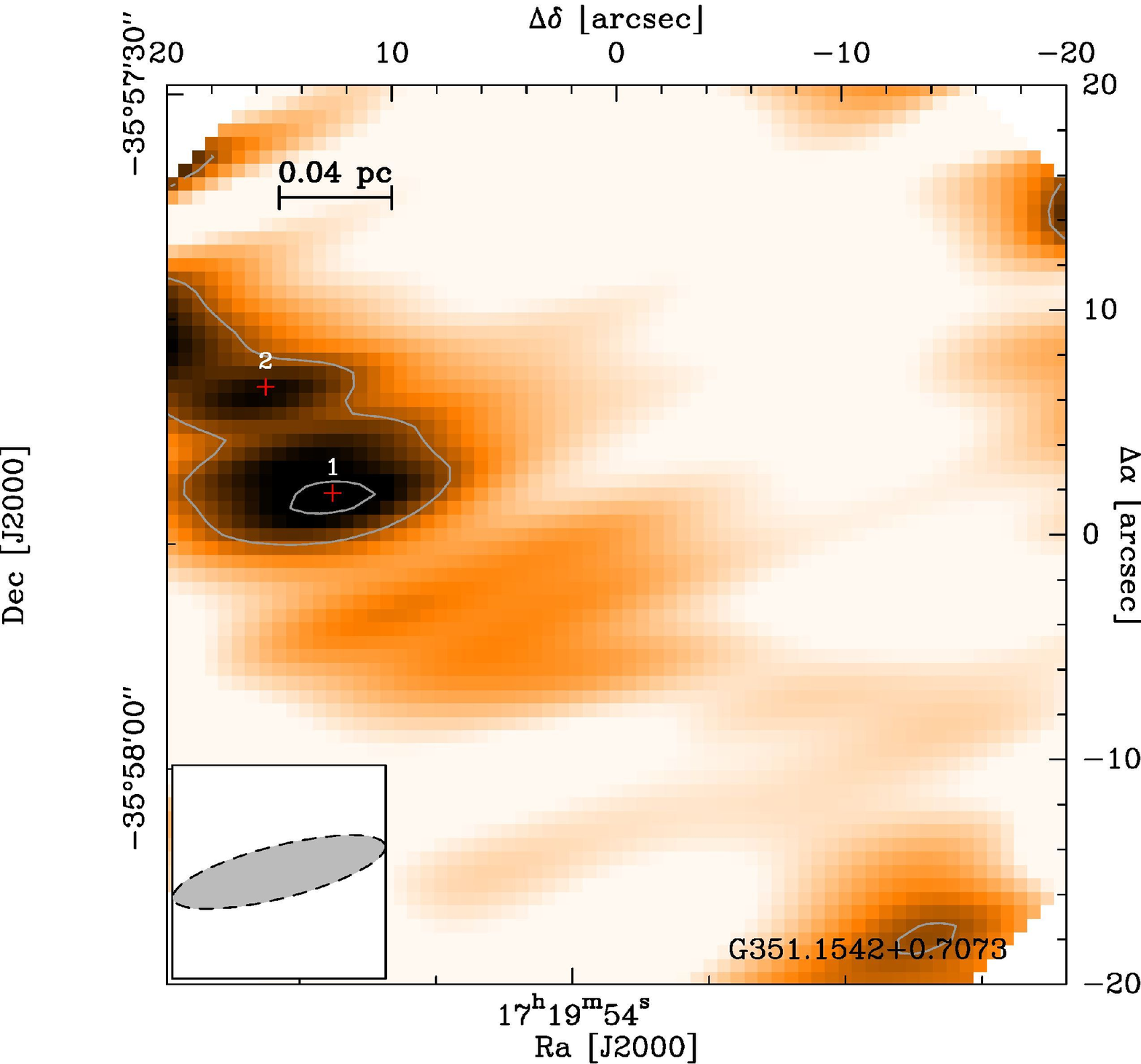}
    \includegraphics[width=0.25\linewidth]{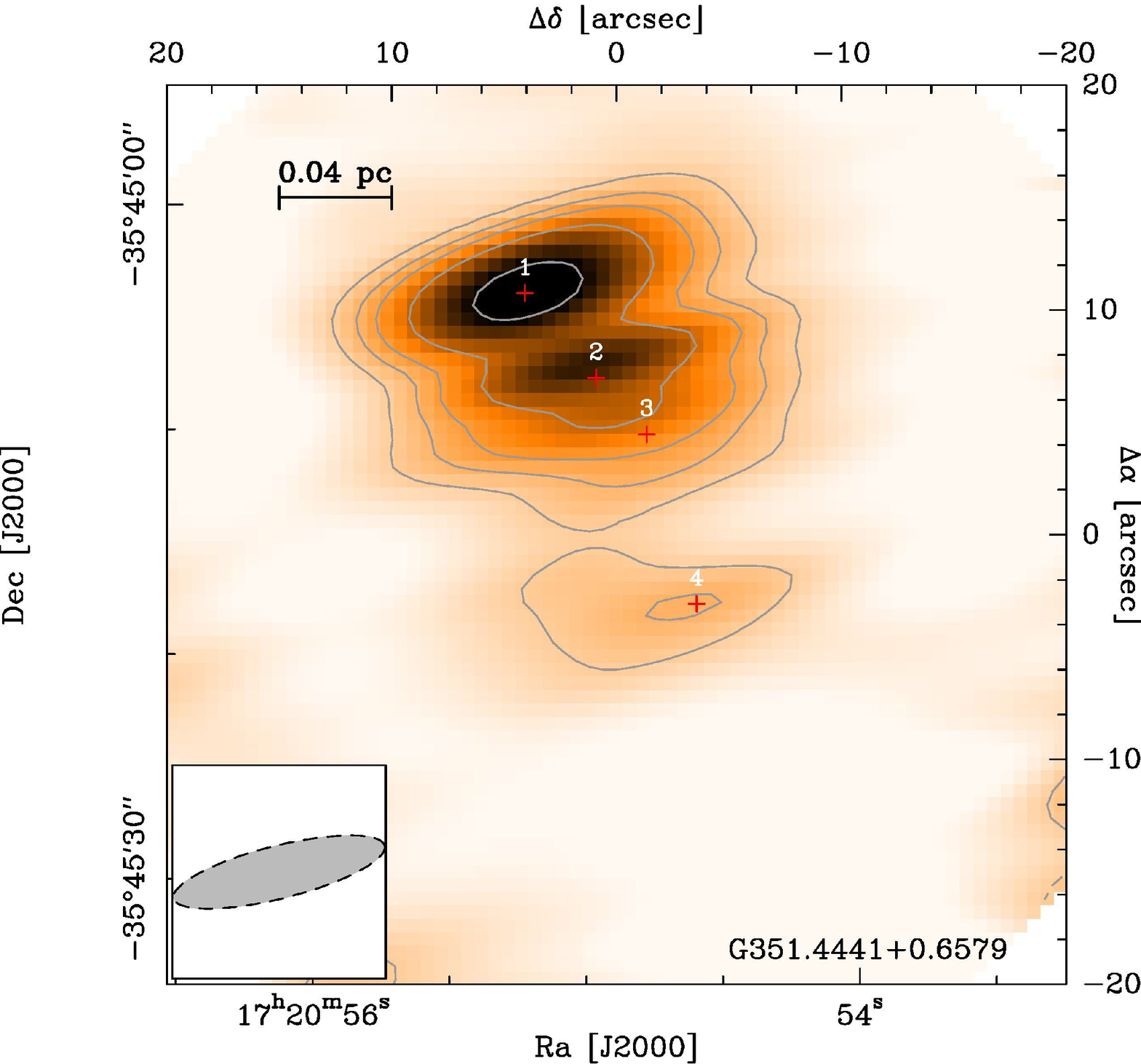}
       \caption{
   		Continued.}
    \end{figure*}
    
       \begin{figure*}[!htpb]
   \centering
   \ContinuedFloat
     \includegraphics[width=0.25\linewidth]{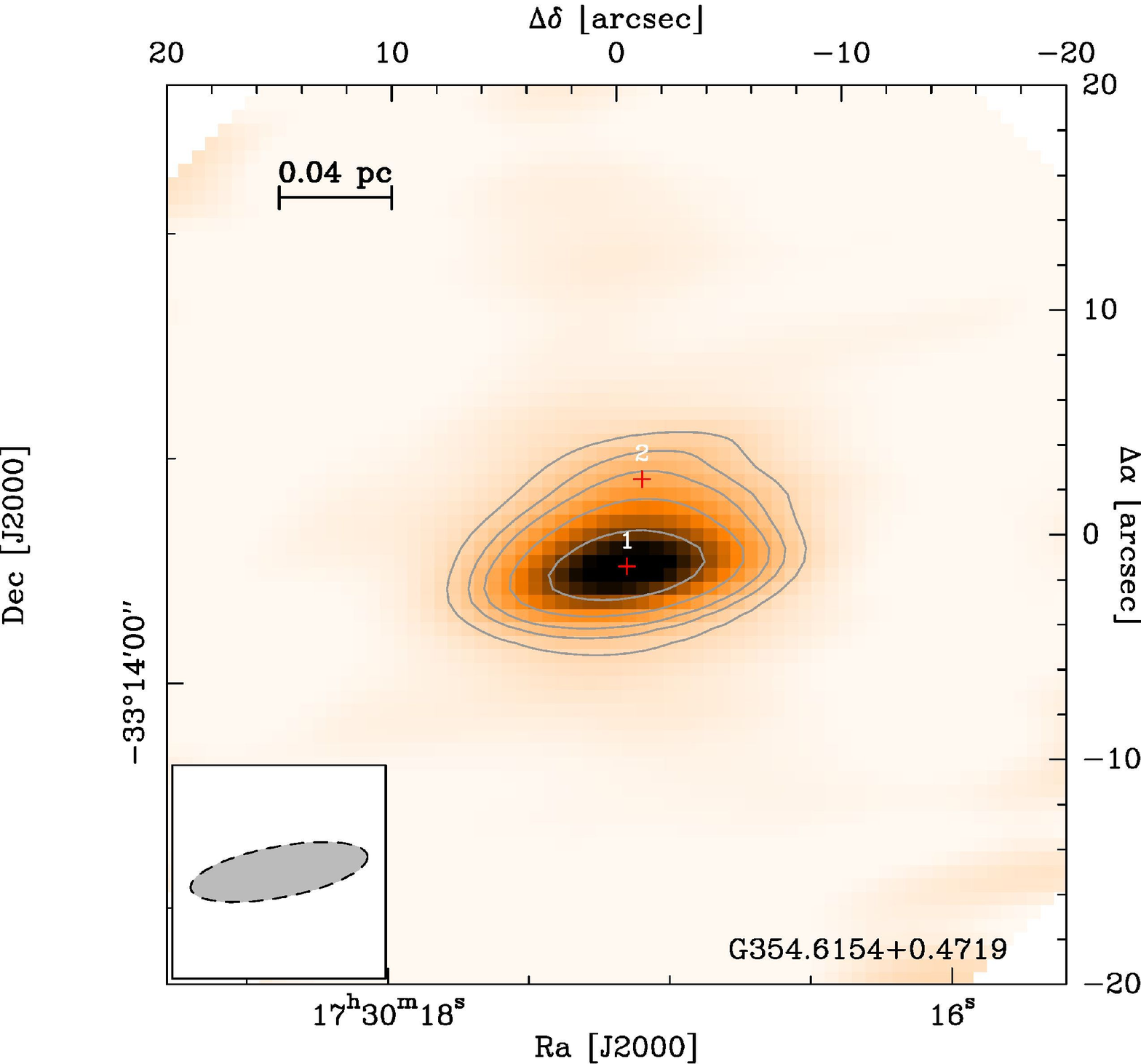}
   \includegraphics[width=0.25\linewidth]{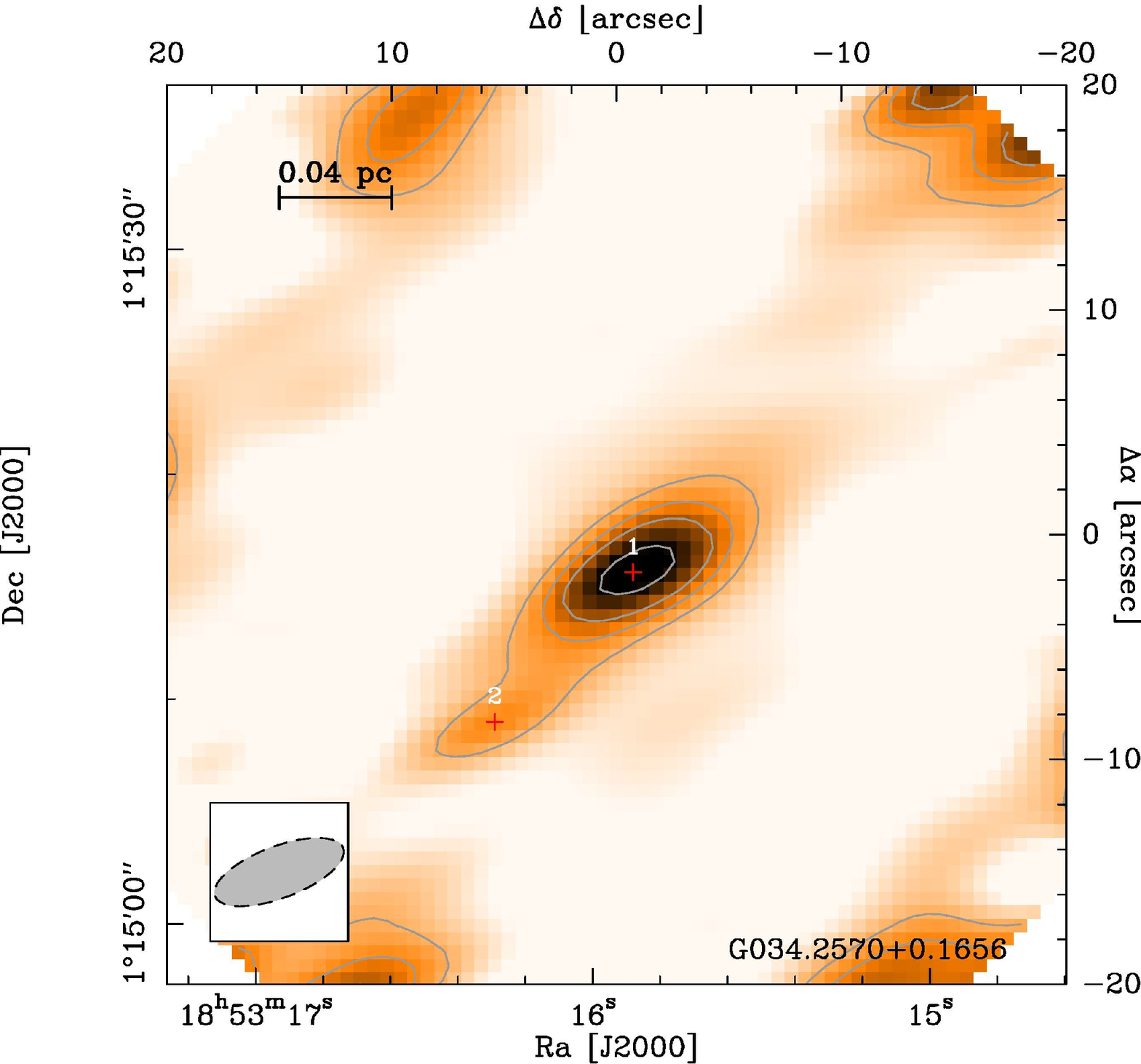}
    \includegraphics[width=0.25\linewidth]{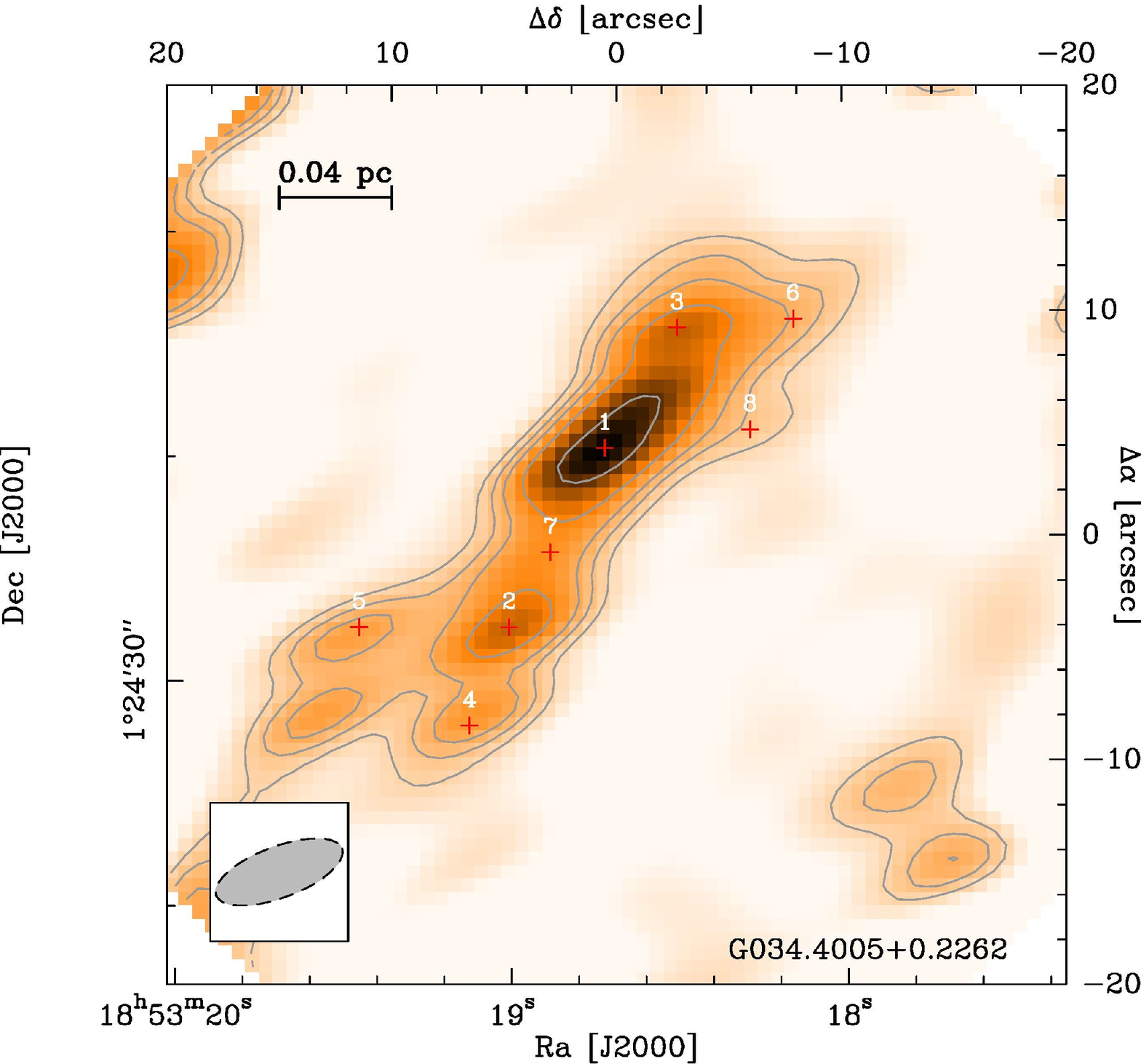}
    \includegraphics[width=0.25\linewidth]{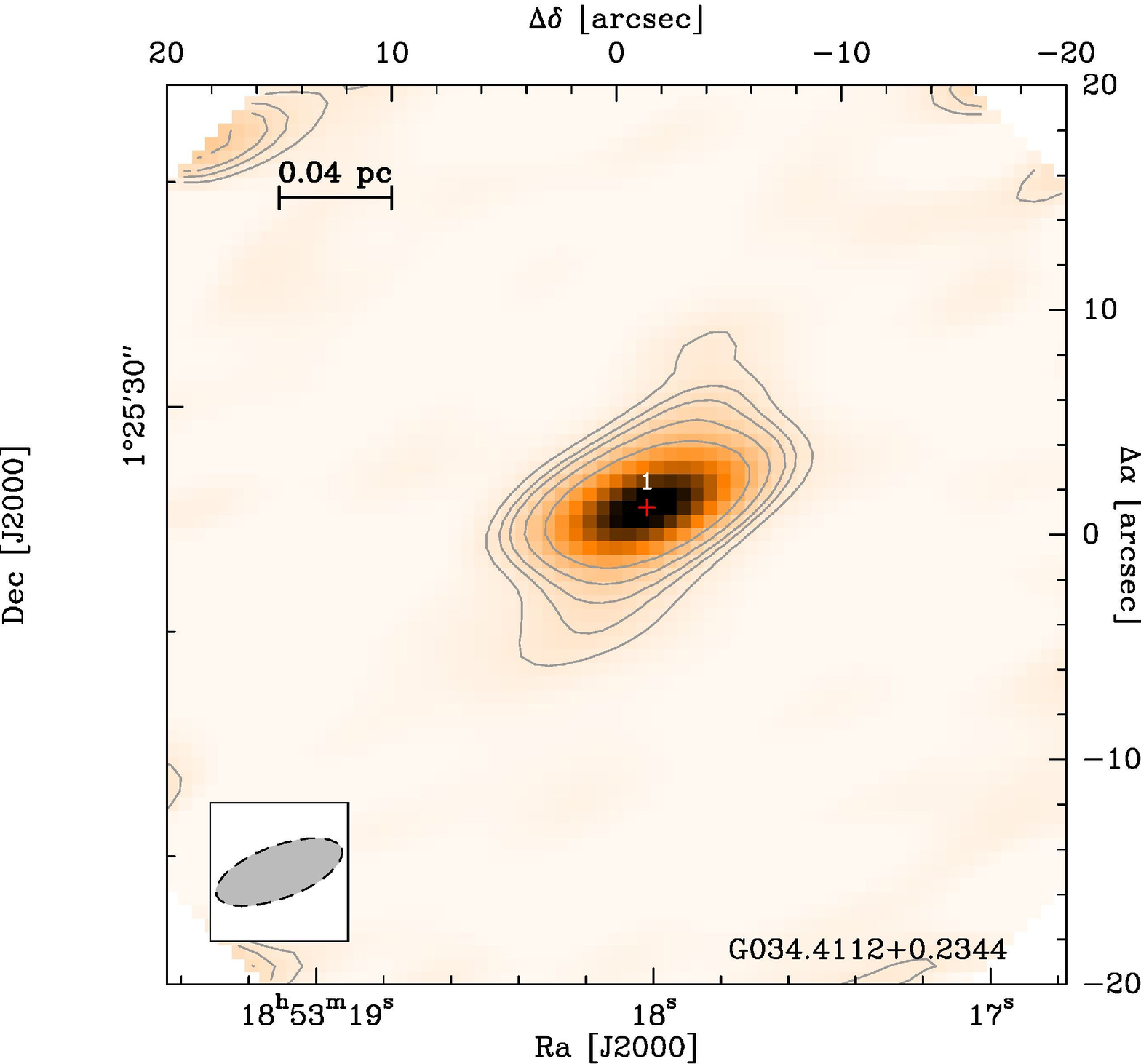}
    \includegraphics[width=0.25\linewidth]{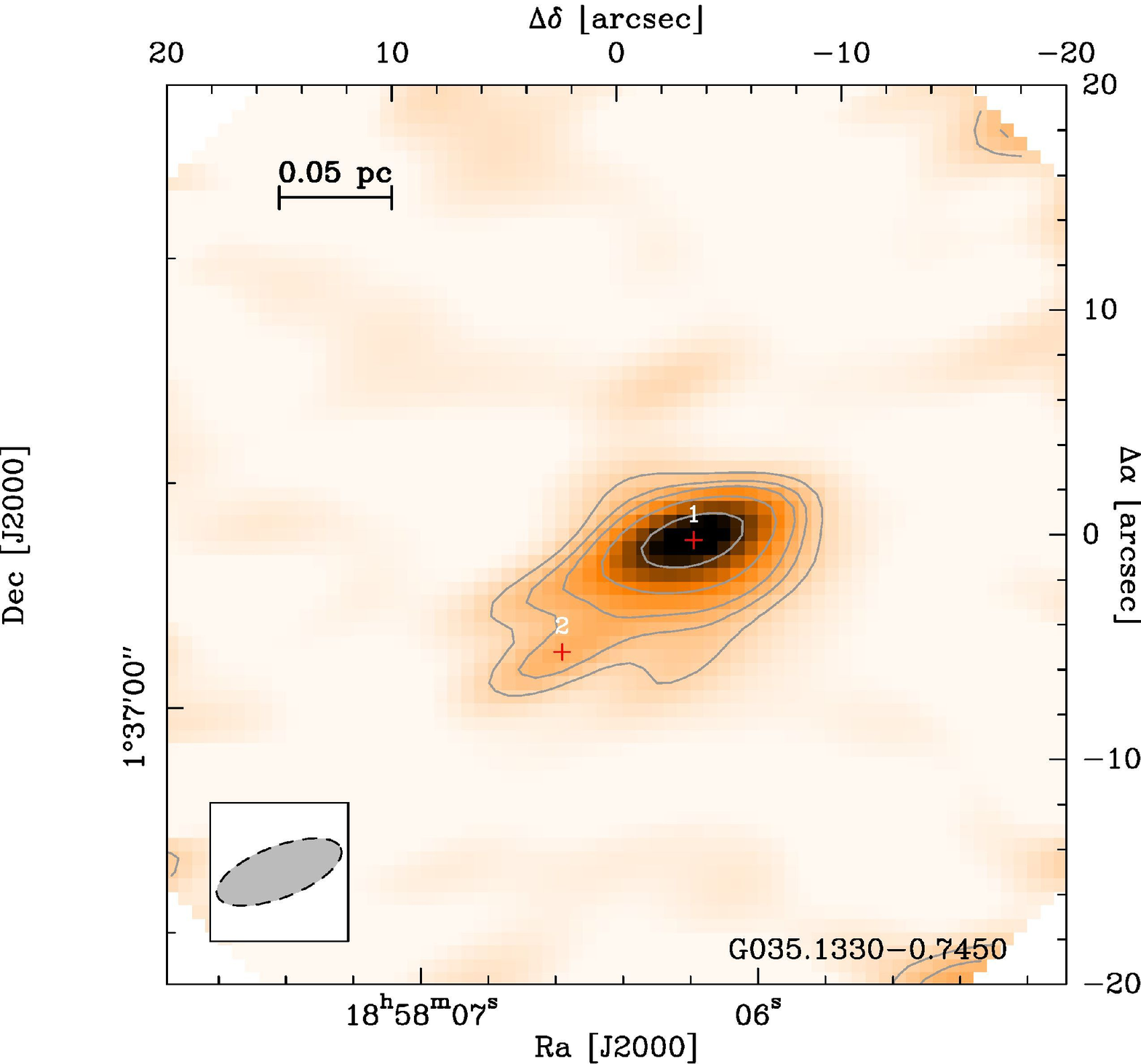}  
   \caption{
   		Continued.}
    \end{figure*}

\end{appendix}

%
%

\end{document}